\documentclass[journal=jcim,manuscript=article]{achemso}
\setkeys{acs}{maxauthors=1000}

\usepackage{soul}
\usepackage{chemformula} %
\usepackage[T1]{fontenc} %
\usepackage{caption}
\usepackage{subcaption}
\usepackage{graphicx}               %
\usepackage{amsmath,amssymb}        %
\usepackage{tabularx,booktabs}
\usepackage{pifont}
\usepackage{todonotes}
\usepackage{xifthen}
\usepackage{mathtools}
\usepackage[version=4]{mhchem}
\usepackage{xspace}
\usepackage{float}
\usepackage{placeins}
\usepackage{bm}

\usepackage{nameref}

\usepackage{xr}
\newcommand{\ourModel}{\texttt{ICEBERG}\xspace}

\newcommand{\scarfModel}{\texttt{SCARF}\xspace}

\newcommand{\ourModelOneShort}{\texttt{Generate}\xspace}
\newcommand{\ourModelTwoShort}{\texttt{Score}\xspace}

\newcommand{\cfmModel}{\texttt{CFM-ID}\xspace}

\newcommand{\nistData}{NIST20\xspace}
\newcommand{\gnpsData}{NPLIB1\xspace}
\usepackage{comment}

\newcommand{\MAGMA}{\texttt{MAGMa}\xspace}
\newcommand{\metfrag}{\texttt{MetFrag}\xspace}
\newcommand{\neimsFFN}{\texttt{NEIMS (FFN)}\xspace}
\newcommand{\neimsGNN}{\texttt{NEIMS (GNN)}\xspace}
\newcommand{\Massformer}{\texttt{MassFormer}\xspace}
\newcommand{\MolMS}{\texttt{3DMolMS}\xspace}
\newcommand{\FixedVocab}{\texttt{FixedVocab}\xspace}
\newcommand{\graffMS}{\texttt{GRAFF-MS}\xspace}

\newcommand{\mol}{\mathcal{M}}
\newcommand{\gnn}[1]{\textsf{GNN}(#1)}
\newcommand{\transformer}[1]{\textsf{Transformer}(#1)}
\newcommand{\mlp}{\textsf{MLP}}

\newcommand{\softmax}{\textsf{Softmax}}

\newcommand{\formEnc}{\textsf{Enc}}
\newcommand{\oneHot}{\textsf{Onehot}}

\newcommand{\molnodes}{N}
\newcommand{\moledges}{E}
\newcommand{\substructsymb}{\mathcal{S}}
\newcommand{\substruct}[1]{\substructsymb^{(#1)}}
\newcommand{\subnodes}[1]{N^{(#1)}}
\newcommand{\subedges}[1]{E^{(#1)}}
\newcommand{\numBroken}{b}

\newcommand{\tree}{\mathcal{T}}
\newcommand{\form}{f}
\newcommand{\treenodes}{\substructsymb}
\newcommand{\treenodesind}[1]{\substruct{#1}}
\newcommand{\formInd}[1]{\form_{#1}}

\newcommand{\treeedges}{\mathcal{E}}
\newcommand{\fragmentation}{\mathcal{F}}
\newcommand{\spec}{\mathcal{Y}}
\newcommand{\predspec}{\hat{\mathcal{Y}}}

\newcommand{\predspecbinned}{\hat{\bm{s}}}
 \newcommand{\specbinned}{\bm{s}}

 \newcommand{\predspecbinnedind}[1]{\hat{s}_{#1}}
 \newcommand{\specbinnedind}[1]{s_{#1}}

\newcommand{\massfunc}{M}
\newcommand{\specsim}{\text{sim}}

\newcommand{\ltwonorm}[1]{\left \lVert #1 \right \rVert_2}

 \newcommand{\codeUrl}{\url{https://github.com/samgoldman97/ms-pred}}

\definecolor{modelOneColor}{HTML}{489DD7} %
\definecolor{modelTwoColor}{HTML}{3D4A9F}

 \newcommand{\titleName}{Generating Molecular Fragmentation Graphs with Autoregressive Neural Networks}

\author{Samuel Goldman}
\affiliation{Computational and Systems Biology, Massachusetts Institute of Technology, 
Cambridge, Massachusetts 02139, USA}

\author{Janet Li}
\affiliation{Harvard College, Harvard University, Cambridge, Massachusetts 02138, USA}

\author{Connor W. Coley}
\affiliation[ChemE]{Department of Chemical Engineering, Massachusetts Institute of Technology, 
Cambridge, Massachusetts 02139, USA}
\alsoaffiliation{Department of Electrical Engineering and Computer Science, Massachusetts Institute of Technology, 
Cambridge, Massachusetts 02139, USA}
\email{* ccoley@mit.edu}

\title{\titleName{}}

\begin{document}

\begin{tocentry}

\centering
\includegraphics[width=1\textwidth]{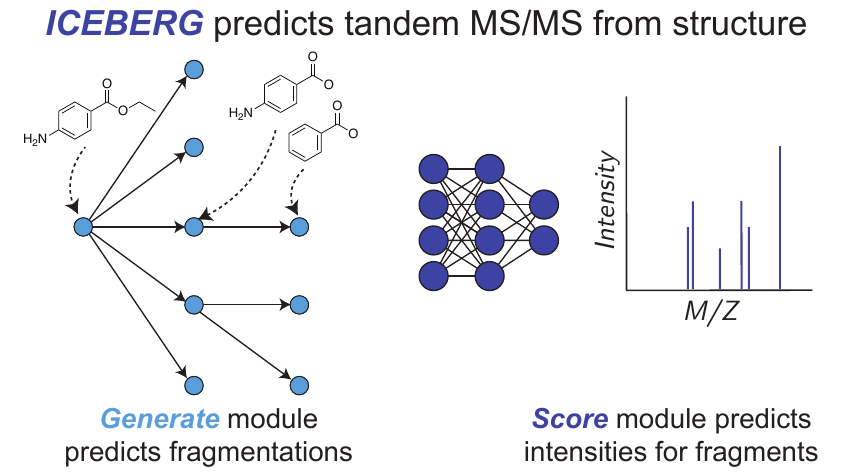}
\label{fig:toc}

\end{tocentry}

\begin{abstract}
The accurate prediction of tandem mass spectra from molecular structures has the potential to unlock new metabolomic discoveries by augmenting the community's libraries of experimental reference standards. Cheminformatic spectrum prediction strategies use a “bond-breaking” framework to iteratively simulate mass spectrum fragmentations, but these methods are (a) slow, due to the need to exhaustively and combinatorially break molecules and (b) inaccurate, as they often rely upon heuristics to predict the intensity of each resulting fragment; neural network alternatives mitigate computational cost but are black-box and not inherently more accurate.  We introduce a physically-grounded neural approach that learns to predict each breakage event and score the most relevant subset of molecular fragments quickly and accurately. We evaluate our model by predicting spectra from both public and private standard libraries, demonstrating that our hybrid approach offers state of the art prediction accuracy, improved metabolite identification from a database of candidates, and higher interpretability when compared to previous breakage methods and black box neural networks. The grounding of our approach in physical fragmentation events shows especially high promise for elucidating natural product molecules with more complex scaffolds.
\end{abstract}


\section{Introduction}
\label{sec:intro}
\begin{figure}
  \centering
  \vspace{-2em}
  \includegraphics[width=0.5 \textwidth]{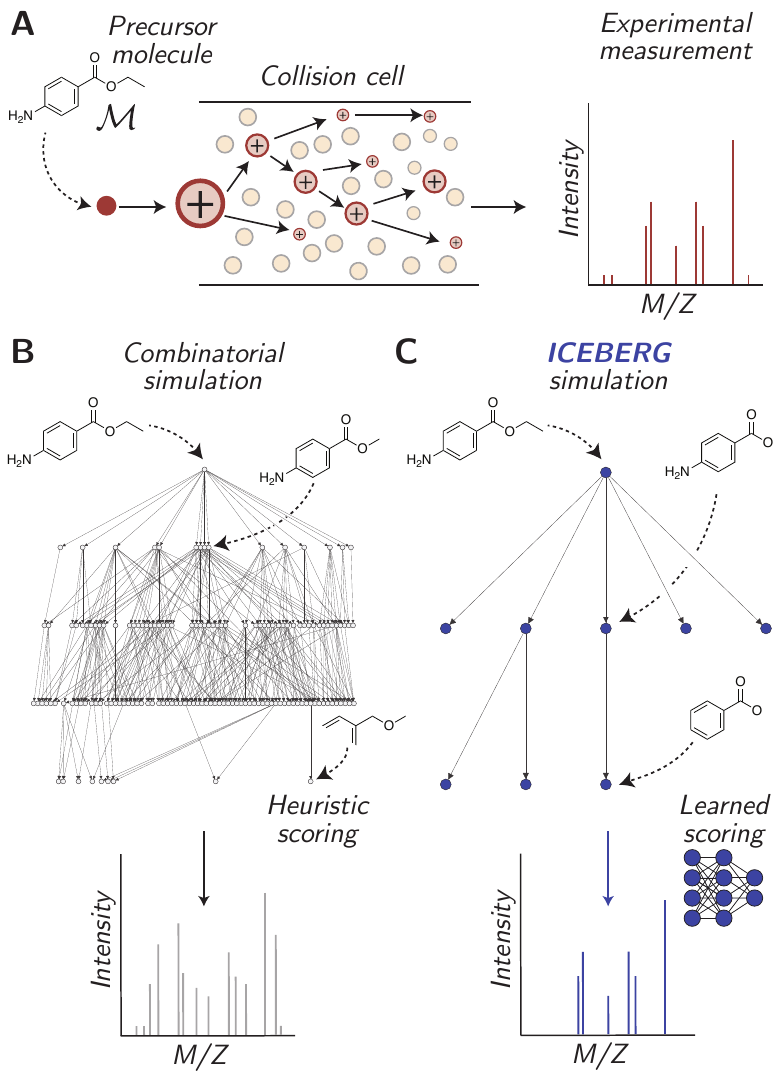}
  \caption{
    ICEBERG enables the prediction of tandem mass spectra by efficiently navigating the space of possible fragmentation events. \textbf{A.} Example experimental mass spectrum. An input molecule, benzocaine, is depicted entering a mass spectrometer collision cell and fragmenting. The observation of the resulting charged fragments results in a characteristic spectrum. \textbf{B.} A combinatorial mass spectrum simulation. The root molecule, benzocaine, is iteratively fragmented by removing atoms or breaking bonds, resulting in a large fragmentation tree. Heuristic rules score nodes in the tree to predict intensities. \textbf{C.} \ourModel spectrum simulation. \ourModel learns to generate only the most relevant substructures. After generating fragments, a neural network module scores the resulting fragments to predict intensities. %
  }
\label{fig:icebergTeaser}
\end{figure}

Identifying unknown molecules in complex metabolomic or environmental samples is of critical importance to biologists\cite{wishart_metabolomics_2019}, forensic scientists\cite{szeremeta_ijms_2021}, and ecologists alike\cite{bundy_environmental_2009}. Tandem mass spectrometry, MS/MS, is the standard analytical chemistry method for analyzing such samples, favored for its speed and sensitivity \cite{neumann_computational_2010}. In brief, MS/MS metabolomics experiments isolate, ionize, and fragment small molecules, resulting in a characteristic spectrum for each where peaks correspond to molecular sub-fragments (Figure \ref{fig:icebergTeaser}A). Importantly, these experiments are high throughput, leading to thousands of detected spectra per single experiment for complex samples such as human serum. 

The most straightforward way to identify an unknown molecule from its fragmentation spectrum is to compare the spectrum to a library of known standards \cite{bittremieux2022critical}. However, spectral libraries only contain on the order of $10^4$ compounds---a drop in the bucket compared to the vast size of biologically-relevant chemical space, oft cited as large as $10^{60}$ \cite{kirkpatrick2004chemical}. Of the many tandem spectra deposited into a large community library, 87\% still cannot be annotated \cite{bittremieux2022critical}. The accurate prediction of mass spectra from molecular structures would enable these libraries to be augmented with hypothetical compounds and significantly advance the utility of mass spectrometry for structural elucidation. This paradigm of comparing unknown spectra to putative spectra is well established in the adjacent field of proteomics due to the ease of predicting protein fragmentations  \cite{frewen2006analysis}.

Because tandem mass spectrometry experiments physically break covalent bonds in a process known as ``collision-induced-dissociation'' (CID) to create fragments, simulating such fragmentation events computationally is a natural strategy for prediction. Tools from the last decade including \metfrag \cite{wolf_silico_2010}, \MAGMA \cite{ridder_automatic_2014}, and \cfmModel \cite{allen_competitive_2015, wang_cfm-id_2021} use fragmentation rules (based on removing atoms or bonds) and local scoring methods to (a) enumerate molecular fragmentation trees and (b) estimate the intensity at each node in the tree with a mix of heuristic rules and statistical learning (Figure \ref{fig:icebergTeaser}B).

However, these combinatorial methods are computationally demanding and often make inaccurate predictions by \emph{overestimating} the possible fragments (Figure \ref{fig:icebergTeaser}B, bottom). 
We recently found \cfmModel to be far less accurate than black-box neural networks \cite{goldman_prefix-tree_2023}, an observation separately confirmed by \citeauthor{murphy2023efficiently} \cite{murphy2023efficiently} Further, current learned fragmentation models are not easily adapted or scaled to new datasets; \citeauthor{murphy2023efficiently} estimate it would take the leading fragmentation approach, \cfmModel \cite{allen_competitive_2015}, approximately three months on a 64-core machine to train on an approximately 300,000 spectrum dataset.

Alternative strategies that utilize black box neural networks to predict MS/MS spectra have been attempted. They encode an input molecule (e.g., as a fingerprint, graph, or 3D structure) and predict either a 1D binned representation of the spectrum \cite{wei_rapid_2019,
zhu_using_2020, young_Massformer_2021, hong20233DMolMS}, or a set of output formulae corresponding to peaks in the spectrum \cite{goldman_prefix-tree_2023, zhu2023rapid, murphy2023efficiently}. While we have demonstrated that predicting chemical formulae provides a fast, accurate, and interpretable alternative to binned representation approaches \cite{goldman_prefix-tree_2023}, the improved accuracy surprisingly did not directly translate to better database retrieval for complex natural product molecules contained within the Global Natural Products Social (GNPS) database  \cite{wang_sharing_2016}. We hypothesized that combining the flexibility of neural networks to learn from experimental MS/MS data in reference libraries with the structural-bias of combinatorial fragmentation approaches could lead to increased prediction performance on complex natural product molecules.

Herein, we introduce a hybrid strategy for simulating molecular fragmentation graphs using neural networks, \emph{Inferring Collision-induced-dissociation by Estimating Breakage Events and Reconstructing their Graphs} (\ourModel). \ourModel is a two-part model that simulates  probable breakage events (\ourModelOneShort) and scores the resulting fragments using a Transformer architecture (\ourModelTwoShort) (Figure \ref{fig:icebergTeaser}C; details in Figure \ref{fig:methods}). Our core computational contribution is to leverage previous exhaustive cheminformatics methods for the same task, specifically \MAGMA \cite{ridder_automatic_2014}, in order to build a %
training dataset, from which our model learns to make fast estimates prioritizing only likely bond breakages. In doing so, we lift \MAGMA and previous bond-breaking approaches into a neural network space with demonstrable benefits in performance. 

We evaluate \ourModel on two  datasets: \gnpsData (GNPS data \cite{wang_sharing_2016} as used to train the CANOPUS model \cite{duhrkop_systematic_2021}) and \nistData \cite{noauthor_tandem_nodate}, which test the model's ability to predict both complex natural products and small organic standard molecules, respectively. We find that \ourModel increases cosine similarity of predicted spectra by 10\%  (0.63 vs. 0.57) compared to a recent state of the art method on \gnpsData data. When used to identify molecules in retrospective retrieval studies, \ourModel leads to a 46\% relative improvement (29\% vs. 20\%) in top 1 retrieval accuracy on a challenging natural product dataset compared to the next best model tested.  \ourModel is fully open-sourced with pretrained weights alongside other existing prediction baseline methods available on GitHub at \codeUrl.

\section{Methods}
\label{sec:methods}

\subsection{Datasets}
\label{sec:methods_datasets}

We train our models on the two datasets, \nistData \cite{noauthor_tandem_nodate} as generated by the National Institute of Standards  and \gnpsData extracted from the GNPS database \cite{wang_sharing_2016} and prepared previously by \citeauthor{duhrkop_systematic_2021}\cite{duhrkop_systematic_2021} and  \citeauthor{Goldman2023annotating}\cite{Goldman2023annotating} For each spectrum in the dataset, we first merge all scans at various collision energies, combine peaks that are within $10^{-4}$ m/z tolerance from each other, renormalize the resulting spectrum by dividing by the maximum observed intensity, and take the square-root of each intensity. We subset the resulting spectrum to keep the top 50 peaks with intensity above $0.003$. This normalization process is identical to our previous work \cite{goldman_prefix-tree_2023} and emphasizes (a) removing peaks that are likely noise and (b) combining various collision energies. We refer the reader to our previous work~\cite{goldman_prefix-tree_2023} for exact details on dataset extraction. 

To further normalize the dataset, for each spectrum, we subtract the mass of the adduct ion from each resulting MS2 peak. Concretely, the precursor molecule is ionized with an adduct ion, for instance, \ch{H}$^+$. In this case, the mass of each peak in the spectrum is shifted by the mass of \ch{H}$^+$ before proceeding further. In doing so, we normalize against different ionizations. While adduct switching is possible, this assumption allows us to make predictions in a more constrained model output space (i.e., models do not need to predict all combinations of potential adduct masses at each predicted fragment) and allows our models to share output representations across different input adduct types (i.e., potentially learning how to predict rarer fragments more accurately). Further, as described below, our models are capable of predicting various hydrogen-shifted peak intensities that can serve to regularize and correct wrongly annotated adduct switching patterns in a small number of cases (e.g., \ce{[M{+}H]+} $\rightarrow$ \ce{[M]+}).We defer more complete incorporation of adduct switching into modeling as a potential direction for future work. We make the simplifying assumption that all peaks are singly charged and use mass and m/z interchangeably.  Ultimately, each spectrum $\spec$ can be considered a set of mass, intensity tuples, $\spec = \{(m_0, y_0), (m_1, y_1), \dots\, (m_{|\spec|}, y_{|\spec|})\}$.

\subsection{Canonical DAG construction}
\label{sec:methods_dag}

We build a custom re-implementation of the \MAGMA algorithm \cite{ridder_automatic_2014} to help create explanatory directed acyclic grpahs (DAGs) for each normalized and adduct-shifted spectrum. 

Given an input molecule $\mol$, \MAGMA iteratively breaks each molecule by removing atoms. Each time an atom is removed, multiple fragments may form, from which we keep all fragments of $>2$ heavy (non-hydrogen) atoms. To prevent combinatorial explosion of DAG nodes, we use a Weisfeiler-Lehman isomorphism test \cite{weisfeiler1968reduction} to generate a unique hash ID of each generated fragment and reject new fragments with hash IDs already observed. When conducting this test, to remain insensitive to \emph{how} this fragment originated, we hash only the atom identities and bonds in the fragment graph, \emph{not the number of hydrogen atoms}. For instance, consider an ethane fragment in which the terminal carbon was originally double-bonded to a single neighboring atom in the precursor molecule compared to an ethane fragment in which the terminal carbon was single-bonded to two adjacent atoms in the original precursor--- our approach applies the same hash ID to both fragments. The chemical formula and hydrogen status for the fragment is randomly selected from the fragments that required the \textit{minimal} number of atom removals. Each fragment corresponds to multiple potential m/z observations due to the allowance for hydrogen shifts equal to the number of broken bonds.

After creating the fragmentation graph for $\mol$, a subset of the fragments are selected to explain each peak in $\spec$, using mass differences of under 20 parts-per-million as the primary filter and the minimal \MAGMA heuristic score as a secondary filter. We include nodes along all paths back to the root molecule for each selected fragment. To prune the DAG to select only the most likely paths to each fragment, we design a greedy heuristic. Starting from the lowest level of the DAG, we iteratively select the parent nodes for inclusion into the final DAG that ``cover'' the highest number of peak-explaining nodes. Finally, the ``neutral loss'' fragments are added into the DAG, as they provide useful training signals for \ourModel \ourModelOneShort to learn when to stop fragmenting each molecule. By using these heuristics, it is likely that we select incorrect fragment annotations for a subset of peaks in our training dataset. Our intuition is that a deep learning model trained on this data will still be able to make accurate and generalizable predictions to aid in small molecule structure elucidation, even if the level of abstraction is not always true or exactly matched to how an expert chemist may annotate certain MS2 peaks; empirical results demonstrate that this is the case.

\subsection{Model details}
\label{sec:methods_model}

We introduce \ourModel to predict spectra from input molecules. At a high level, our model is split into two components where the first model generates molecular substructure (fragment) candidates and the second model predicts intensities and hydrogen rearrangements for each generated fragment.  In doing so, we simplify the complexity of both models: the first model outputs a set of likely substructures without needing to handle rearrangements and the second model predicts potential rearrangement peak shifts and intensities for the constrained set of potential fragments. We describe the methodology for both models separately below. 

\paragraph{DAG generation prediction} 
We train a model to recursively select the atoms around which to break bonds. The model begins by making predictions on the input molecule. We cheminformatically generate the resulting substructures and iteratively apply our model to each. To train this model, we utilize the ``ground truth DAG'' as described above and train \ourModel \ourModelOneShort to reconstruct the DAG from an input molecule and input adduct type. We define the root fragment in the DAG as the input molecule $\treenodesind{0}$. We can generally define the predicted fragmentation probability for the bonds around the $j^{th}$ atom node in the $i^{th}$ generated fragment in the DAG, $\treenodesind{i}$:

\begin{equation}
p\left(\fragmentation[\treenodesind{i}_j] | \treenodesind{i}, \mol, C\right) = \textcolor{modelOneColor}{g_{\theta}^{\ourModelOneShort}}(\mol,\treenodesind{i}, C)_j  
\end{equation}

As can be seen in this equation, predictions are mode atom-wise (i.e., at every single atom for each fragment in the DAG). To make this atom-wise prediction, we encode information about the root molecule, fragment molecule, their difference, their respective chemical formulae, the adduct, and the number of bonds that were broken between the root molecule and fragment. To embed the root molecule, we utilize a gated graph neural network \cite{Li2015-fc},  $\gnn{\mol}$, where either average or weighted summations are used to pool embeddings across atoms (specified by a hyperparameter). We utilize the same network to learn representations of the fragment,  $\gnn{\treenodesind{i}}$ and define $\gnn{\treenodesind{i}}_j$ as the graph neural network-derived embedding of fragment $i$ at the $j^{th}$ atom prior to pooling operation. For all graph neural networks, a one-hot encoding of the adduct type is also added as atom-wise features alongside the bond types and atom types. We define the chemical formula $\form$ for each DAG fragment and specify an encoding, $\formEnc$, using the Fourier feature scheme defined in \cite{goldman_prefix-tree_2023}. We encode the root and $i^{th}$ node of the fragmentation DAG  as $\formEnc(\formInd{0})$ and  $\formEnc(\formInd{i})$, respectively. Lastly, we define a one hot vector for the number of bonds broken, $\numBroken$. 

All the encodings described above are concatenated together and a shallow multilayer perceptron (MLP) ending with a sigmoid function is utilized to predict binary probabilities of fragmentation at each atom. 

\begin{equation}
\begin{split}
p\left(\fragmentation[\treenodesind{i}_j] | \treenodesind{i}, \mol, C\right) = \mlp \Big([&\gnn{\mol},\gnn{\mol} - \gnn{\treenodesind{i}}, \\
&\gnn{\treenodesind{i}}_j, \oneHot(\numBroken),\formEnc(\formInd{i}) ,\formEnc(\formInd{0} - \formInd{i})\big]\Big)
\end{split}
\end{equation}

The model is trained to maximize the probability of generating the DAG by minimizing the binary cross entropy loss over each atom for every fragment in an observed spectrum.

\paragraph{DAG intensity prediction} The trained \ourModelOneShort module is used to generate DAGs for each input molecule in the training set by iteratively fragmenting the input molecule, with the probability of each fragment computed autoregressively. Once this fragment DAG is created, the \ourModel \ourModelTwoShort module learns to predict intensities at each fragment. Because we deferred rearrangement predictions within \ourModel \ourModelOneShort, we design \ourModel \ourModelTwoShort to simultaneously predict intensities at each generated fragment mass \emph{and} at a range of masses representing potential hydrogen additions or removals.

We define the node indices for an ordering from each fragment $\treenodesind{i}$ back to the root node through its highest likelihood path $\pi[i]$, where $\pi[i,j]$ defines the $j^{th}$ node on this factorization path.

\begin{equation}
p(\treenodesind{i} | \mol, C) =  p(\treenodesind{i} |  \treenodesind{\pi([i,1]}, \mol, C) \prod_{j =1}^{ |\pi[i]|} p(\treenodesind{\pi[i,j]} | \treenodesind{\pi[i, j+1]}, \mol, C) 
\end{equation}

At each step, we maintain only the top $100$ most likely fragments in the DAG as a practical consideration until reaching the maximum possible fragmentation depth. To further reduce complexity in the inference step, we maintain the highest scoring isomer from the DAG. This resulting set of fragments is featurized and passed to a Set Transformer module to generate output values at each fragment. Following the notation from the generative model, we featurize each individual fragment with a shallow MLP to generate hidden representations, $h_i$:

\begin{equation}
\begin{split}
h_i = \mlp \Big([&\gnn{\mol},\gnn{\mol} - \gnn{\treenodesind{i}}, \gnn{\treenodesind{i}}, \\
&\oneHot(\numBroken),\formEnc(\formInd{i}) ,\formEnc(\formInd{0} - \formInd{i})\big]\Big)
\end{split}
\end{equation}

These are subsequently jointly embedded with a Transformer module and used to predict unnormalized intensity weights at each possible hydrogen shift $\delta$ alongside an attention weight $\alpha$ to determine how heavily to weight each prediction for its specified hydrogen shift. To compute the attention weight, we take a softmax over all prediction indices that fall into the same intensity bin (0.1 resolution), $\massfunc(i, \delta)$:

\begin{equation}
\hat{y}^{(i)}_{\delta} = \mlp_{inten}  \Big(\transformer{h_0, h_1,h_2,\dots,h_{|\tree|}}_i\Big)_\delta,
\end{equation}

\begin{equation}
\alpha^{(i)}_{\delta} = \softmax_{k \in \massfunc(i, \delta)} \left(\mlp_{attn} \left( \transformer{h_0, h_1,h_2,\dots,h_{|\tree|}}_k\right)\right)_{i, \delta}
\end{equation}

The final intensity prediction for the bin at mass $m$ is then a then a weighted sum over all predictions that fall within this mass bin followed by a sigmoid activation function:

\begin{equation}
    \hat{y}_m = \sigma \Big(\sum_{i}  \sum_\delta \alpha^{(i)}_{\delta} \hat{y}^{(i)}_{\delta} \mathcal{I}[\massfunc(i,\delta) = m] \Big)
\end{equation}

The model is trained to maximize the cosine similarity between the predicted spectrum and ground truth spectrum. While \ourModel \ourModelOneShort is trained to generate the annotated molecular DAG without any binning assumptions, \ourModel \ourModelTwoShort is trained to directly maximize similarity to a vectorized, binned version of the original spectrum to utilize the same loss function as our baseline methods as described below.

\paragraph{Model training} All models are implemented and trained using Pytorch Lightning \cite{Falcon_PyTorch_Lightning_2019}, the Adam optimizer \cite{kingma2014adam}, and the 
 DGL library \cite{Wang2019-ga}. Ray\cite{liaw2018tune} is used to complete hyperparameter optimizations over all models and baselines. 
Models are trained on a single RTX A5000 NVIDIA GPU (CUDA Version 11.6) in under $3$ hours for each module. A complete list of hyperparameters and their definition can be found in the Supporting Information.

\subsection{Baselines}
\label{sec:methods_baselines}

A key component of this work is to extend our robust comparison to previous and contemporary methods \cite{goldman_prefix-tree_2023}. To conduct this benchmarking and specifically emphasize methodological differences rather than data differences, we port and modify code from a number of previous methods including NEIMS \cite{wei_rapid_2019}, NEIMS with a graph neural network \cite{zhu_using_2020} \MolMS \cite{hong20233DMolMS}, SCARF \cite{goldman_prefix-tree_2023}, a modified version of \graffMS we refer to as \FixedVocab \cite{murphy2023efficiently}, \Massformer \cite{young_Massformer_2021}, and \cfmModel \cite{wang_cfm-id_2021} into a single GitHub repository. 

Following our previous approach in \cite{goldman_prefix-tree_2023}, we emphasize conditioning on the same experimental settings of adduct type \emph{across methods} for fair comparison, excluding collision energies and instrument types as extensions for future work. We rigorously hyperparameter optimize each method for our data regime and train each model on data splits with the exception of \cfmModel for which retraining is not feasible \cite{goldman_prefix-tree_2023, young_Massformer_2021, murphy2023efficiently}.

All model predictions are transformed into binned representations for fair evaluation at a bin resolution of $0.1$ from mass $0$ to $1,500$ Da. Further details are included in the Supporting Information. %

\subsection{Spectral similarity evaluations}

Both \ourModel and the presented baseline models are trained to maximize the spectral similarity between each predicted spectrum and the true spectrum, as this is well aligned with the retrieval settings and has been used in previous studies. While there are many different variants of cosine similarity calculations~\cite{demuth2004spectral,huber_spec2vec_2021, li2021spectral}, we specifically compute a vectorized cosine similarity for any predicted spectrum, $\spec$ and true spectrum, $\predspec$. We denote each vectorized spectrum and predicted spectrum as $\specbinned$ and $\predspecbinned$ respectively, in which all intensities are binned into ranges of 0.1 Da from 0 to 1,500 Da (a \emph{max} aggregation function is used to resolve any peaks within the same bin). The vectorized cosine similarity is then defined as:

\begin{equation}
    \specsim(\specbinned, \predspecbinned) = \frac{\specbinned^\top\predspecbinned}{\ltwonorm{\specbinned} \ltwonorm{\predspecbinned}} = \frac{\sum_{i=1}^{|\specbinned|} \specbinnedind{i} \predspecbinnedind{i}}{\sqrt{\sum_{j=1}^{|\specbinned|} \specbinnedind{j}^2} \sqrt{\sum_{j=1}^{|\specbinned|} \predspecbinnedind{j}^2}}
\end{equation}

While this distance metric forces a conversion of exact mass fragment predictions into binned values for \ourModel predictions, it ensures that we avoid unfairly penalizing the binned spectrum models to which we compare our model. Such baseline models are incapable of predicting exact mass spectra outputs and can only be trained on this variant of the cosine similarity~\cite{wei_rapid_2019,
zhu_using_2020, young_Massformer_2021, hong20233DMolMS}.

\section{Results}

\subsection{ICEBERG is trained as a two-stage generative and scoring model}
\label{sec:res_learning_gen}

\begin{figure}[t]
  \centering
  \includegraphics[width=\textwidth]{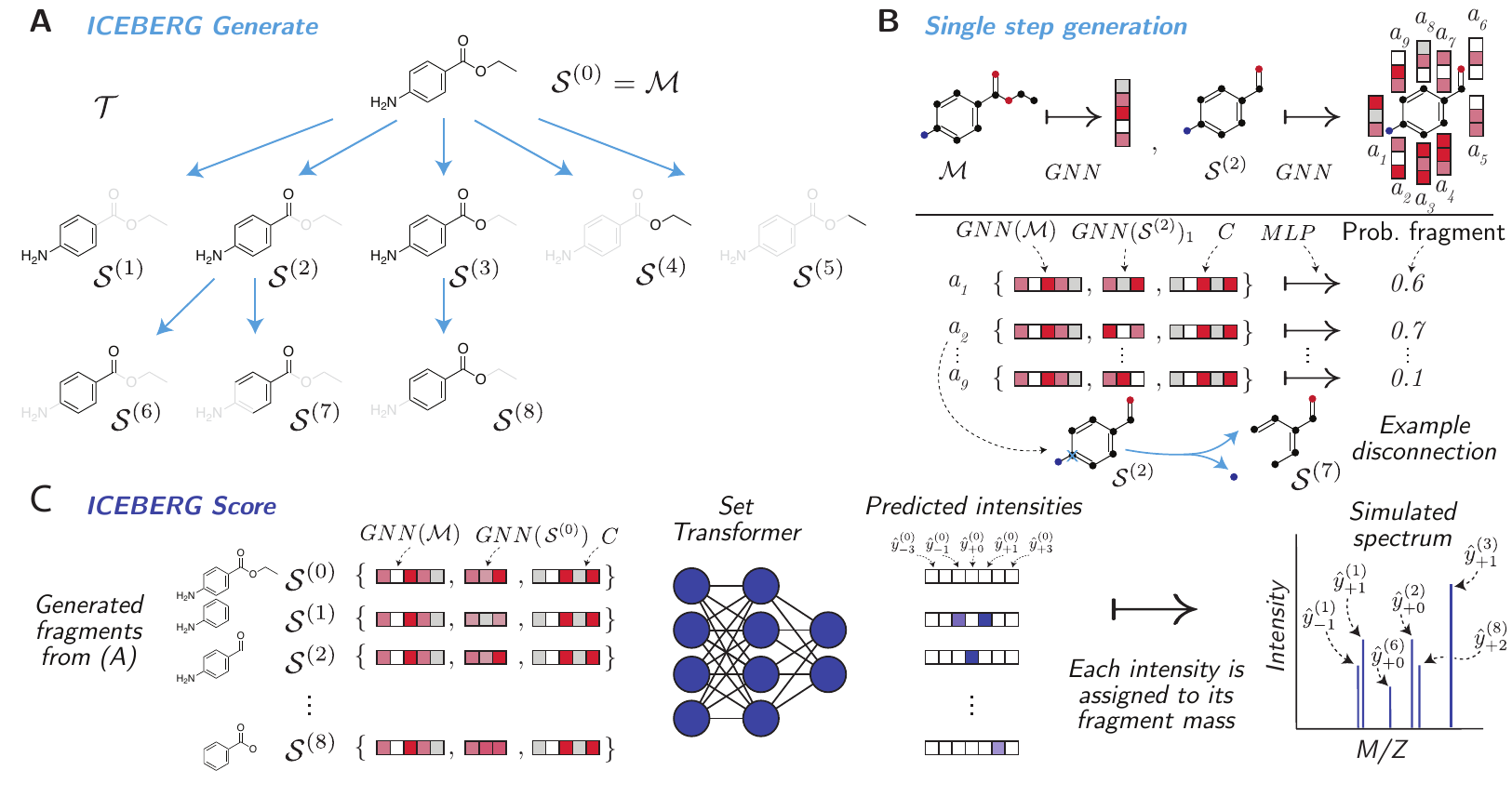}
  \caption{Overview of \ourModel. \textbf{A.} The target fragmentation directed acyclic graph (DAG) for an example molecule $\mol$, benzocaine. Fragments are colored in black with missing substructures in gray.
  \textbf{B.}  Example illustration for the generative process at a single step in the DAG generation predicting subfragments of $\treenodesind{2}$. The root molecule $\mol$, fragment of interest $\treenodesind{2}$, and context vector $C$ are encoded and used to predict fragment probabilities at each atom of the fragment of interest. A sample disconnection is shown at atom $a_2$, resulting in fragment $\treenodesind{7}$.  \textbf{C.}  \ourModel \ourModelTwoShort module. Fragments generated from \textbf{A} are encoded alongside the root molecule. A Set Transformer module predicts intensities for each fragment, allowing mass changes corresponding to the loss or gain of hydrogen atoms, resulting in the final predicted mass spectrum.}
  \label{fig:methods}
\end{figure}

\paragraph{Learning to generate likely substructures.} \ourModel simulates a mass spectrum by generating the substructure fragments from an initial molecule that are most likely to be generated by collision induced dissociation and subsequently measured in the mass spectrometer. We define an input molecule $\mol$ (benzocaine example shown in Figure \ref{fig:methods}A) and its observed spectrum $\spec$, which is a set of intensities at various mass-to-charge values (m/z), termed peaks. Each peak represents one or more observed molecular fragment. 

A core question is then \emph{how to generate the set of potential fragments}. These fragments can be sampled from the many possible substructure options, $\substruct{i} \in (\subnodes{i}, \subedges{i}) \subseteq \mol$, where the set of nodes and edges in substructures are subsets of the atoms and bonds in the original molecule, $\mol \in (\molnodes, \moledges)$. Most often, this sampling is accomplished by iteratively and exhaustively removing edges or atoms from the molecular graph, creating a fragmentation graph $\tree \in (\treenodes, \treeedges)$, where all the nodes in this graph are themselves substructures of the original molecule $\treenodes = \{\substruct{0}, \substruct{1}, \dots \substruct{|\tree|}\}$~\cite{allen_competitive_2015, wolf_silico_2010,  ridder_automatic_2014} (Figure \ref{fig:icebergTeaser}b). However, such a combinatorial approach leads to \emph{thousands} of molecular fragments, making this procedure slow and complicating the second step of estimating intensity values for all enumerated fragments.

We eschew combinatorial generation and instead leverage a graph neural network to parameterize breakage events of the molecule, defining the \ourModelOneShort module of \ourModel (Figure \ref{fig:methods}A,B).  \ourModelOneShort predicts the fragmentation graph iteratively, beginning with just the root of the graph $\treenodesind{0} = \mol$, borrowing ideas from autoregressive tree generation \cite{bradshaw_barking_2020, gao_amortized_2021}. At each step in iterative expansion, the model  $\textcolor{modelOneColor}{g_{\theta}^{\ourModelOneShort}}$ assigns a probability of fragmentation to each atom $j$ in the current substructure fragment $\treenodesind{i}$, $p(F[\treenodesind{i}_j])$. %
Learned atom embeddings are concatenated alongside embeddings of the root molecule and a context vector $C$ containing metadata such as the ionization adduct type in order to make this prediction. %
An illustrative example can be seen for fragment $\treenodesind{2}$  in Figure \ref{fig:methods}B. 
Atom $a_2$ has the highest predicted probability, so this atom is then removed from the graph, leading to the subsequent child node $\treenodesind{7}$ (Figure \ref{fig:methods}B). Importantly, the number of child fragments is determined by how many disjoint molecular graphs form upon removal of the $j^{th}$ node from the molecular graph; in this example, fragments $\treenodesind{1}$ and  $\treenodesind{4}$ originate from the same fragmentation event of $\treenodesind{0}$ (Figure \ref{fig:methods}A). 

In this way, \ourModel predicts breakages at the level of each \emph{atom}, following the convention of \MAGMA \cite{ridder_automatic_2014} rather than each \emph{bond} as is the convention with \cfmModel \cite{allen_competitive_2015}. We strategically use this abstraction to ensure that  all fragmentation events lead to changes in heavy-atom composition. We acknowledge that this formulation does not currently allow for the prediction of skeletal rearrangements and recombinations, which might further improve the model's ability to explain fragmentation spectra. We refer the reader to \nameref{sec:methods_model} for a full description of the model  $\textcolor{modelOneColor}{g_{\theta}^{\ourModelOneShort}}(\mol,\treenodesind{i}, C)_j$,
 graph neural network architectures, and context vector inputs.
 
 While this defines a neural network for generation, we must also specify an algorithm for how to \emph{train} this network. %
 Spectral library datasets contain only molecule and spectrum pairs, but not the directed acyclic graph (DAG) $\tree$ of the molecule's substructures that generated the spectrum. We infer an explanatory substructure identity of each peak for model training by leveraging previous combinatorial enumeration methods, specifically \MAGMA \cite{ridder_automatic_2014}. For each training molecule and spectrum pair, $(\mol, \spec)$, we modify \MAGMA to enumerate all substructures of $\mol$ up to a depth of $3$ sequential fragmentation events. We filter enumerated structures to include only those  with m/z values appearing in the final spectrum, thereby defining a dataset suitable for training \ourModel \ourModelOneShort (see \nameref{sec:methods_dag}). As a result, each paired example $(\mol, \spec)$, in the training dataset is labeled with an estimated fragmentation DAG. \ourModelOneShort learns from these DAGs to generate only the most relevant and probable substructures for a molecule of interest (see \nameref{sec:methods_model}).

\paragraph{Predicting substructure intensities.} After generating a set of potential substructure fragments, we employ a second module, \ourModel \ourModelTwoShort, to predict their intensities (Figure \ref{fig:methods}C). Importantly, this design decision enables our models to consider two important physical phenomena: (i)  neutral losses and (ii) mass shifts due to hydrogen rearrangements and isotope effects.

Because we elect to fragment molecules at the level of atoms (see \nameref{sec:methods_model}), multiple substructures can result from a single fragmentation event. In physical experiments, not all of these substructure fragments will be observed; when fragmentation events occur in the collision cell, one fragment often retains the charge of the parent while the other is uncharged and therefore undetected, termed a ``neutral loss''. By deferring prediction of intensities to a second module, \ourModelOneShort needs not predict or track whether structures are ionized, greatly reducing the complexity of the fragmentation DAG.

In addition to the occurrence of neutral losses, molecules often undergo complex rearrangements in the collision cell, leading to bond order promotions or reductions (e.g., spurious formation of double bonds when a single bond breaks to maintain valence), the most classic of which is the McLafferty rearrangement \cite{demarque2016fragmentation, mclafferty1959mass}. While other approaches attempt to model and estimate where these rearrangements occur using hand-crafted rules \cite{allen_competitive_2015}, we instead adopt the framework of \citeauthor{ridder_automatic_2014} \cite{ridder_automatic_2014} to consider hydrogen tolerances. That is, for each generated molecular substructure $\treenodesind{i}$ we consider the possibility that this fragment is observed not only at its mass, but also at masses shifted by discrete hydrogen masses, $\pm \delta \ch{H}$. This design choice also  simplifies \ourModelOneShort by deferring specification of hydrogen counts to the second model.  In addition to accounting for a mass shift of 1 hydrogen, such flexibility also allows the model to predict the common M+1 isotopes for carbon- and nitrogen- containing compounds.

Mathematically, we define a neural network, $\textcolor{modelTwoColor}{g_{\theta}^{\ourModelTwoShort}}$  that predicts multiple intensities for each fragment $\hat{y}_{\delta}^{(i)}$ corresponding to different hydrogen shifts, $\delta$:
\begin{equation}
  \hat{y}_{\delta}^{(i)} = \textcolor{modelTwoColor}{g_{\theta}^{\ourModelTwoShort}}(\mol,\treenodesind{i}, \tree, C)_\delta
\end{equation}
In practice, we predict up to $13$ intensities at each fragment (i.e., $\{+0\ch{H}, \pm1\ch{H}, \dots, \pm6\ch{H}\}$). For each individual subfragment, the tolerance is further restricted to the number of bonds broken, most often less than $6$. We then take the masses of all fragments, perturb them by the corresponding hydrogen or isotope shifts, and aggregate them into a set of unique m/z peaks by summing the intensities of perturbed fragments with the same m/z value.

To consider all fragments simultaneously in a permutation-invariant manner, $\textcolor{modelTwoColor}{g_{\theta}^{\ourModelTwoShort}}$ is parameterized as a Set Transformer  network \cite{Vaswani2017-se, lee2019set}. %
We train this second module to maximize the cosine similarity between the ground truth spectrum and the predicted spectrum after converting the set of substructures and intensities to m/z peaks. 

\begin{figure}[ht]
  \centering
  \includegraphics[width=\textwidth]{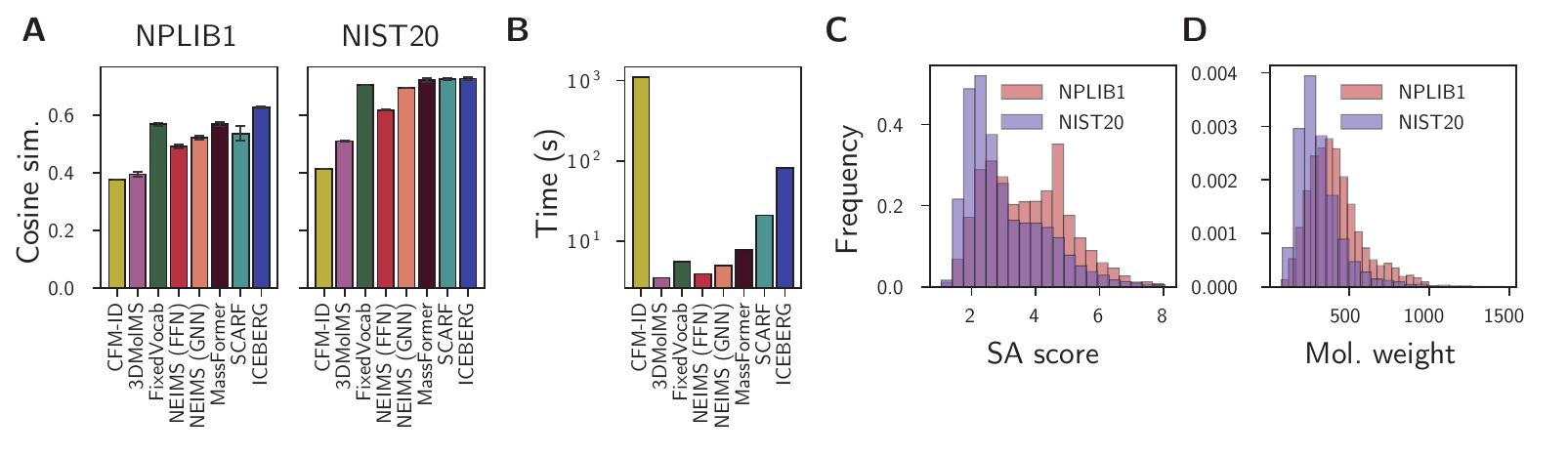}
  \caption{\ourModel predictions are highly accurate. \textbf{A.} Cosine similarities to true spectra on \gnpsData (left) and \nistData respectively (right) for \cfmModel \cite{allen_competitive_2015}, \MolMS \cite{hong20233DMolMS}, \FixedVocab \cite{murphy2023efficiently}, \neimsFFN \cite{wei_rapid_2019}, \neimsGNN \cite{zhu_using_2020}, \Massformer \cite{young_Massformer_2021}, \scarfModel \cite{goldman_prefix-tree_2023}, and \ourModel. Error bars are computed as 1.96 times the standard error of the mean across three random seeds on a single test set split. \textbf{B.} Time required to predict spectra for 100 molecules randomly sampled from \nistData on a single CPU, including the time to load models into memory. \textbf{C,D.} Comparison of \gnpsData and \nistData molecules in terms of synthetic accessibility (SA) score \cite{ertl2009estimation} and molecular weight (Mol. weight).}
  \label{fig:cosine_res}
\end{figure}

At test time, we generate the top 100 most likely fragments from \ourModel \ourModelOneShort and predict intensities for these fragments and their possible hydrogen shifts using \ourModel \ourModelTwoShort. We find this tree size allows our model to consider sufficiently many potential fragments while maintaining a speed advantage over previous fragmentation approaches.

\subsection{\ourModel enables highly accurate spectrum prediction}
\label{sec:spec_results}

\begin{figure}[ht]
  \centering
  \includegraphics[width=\textwidth]{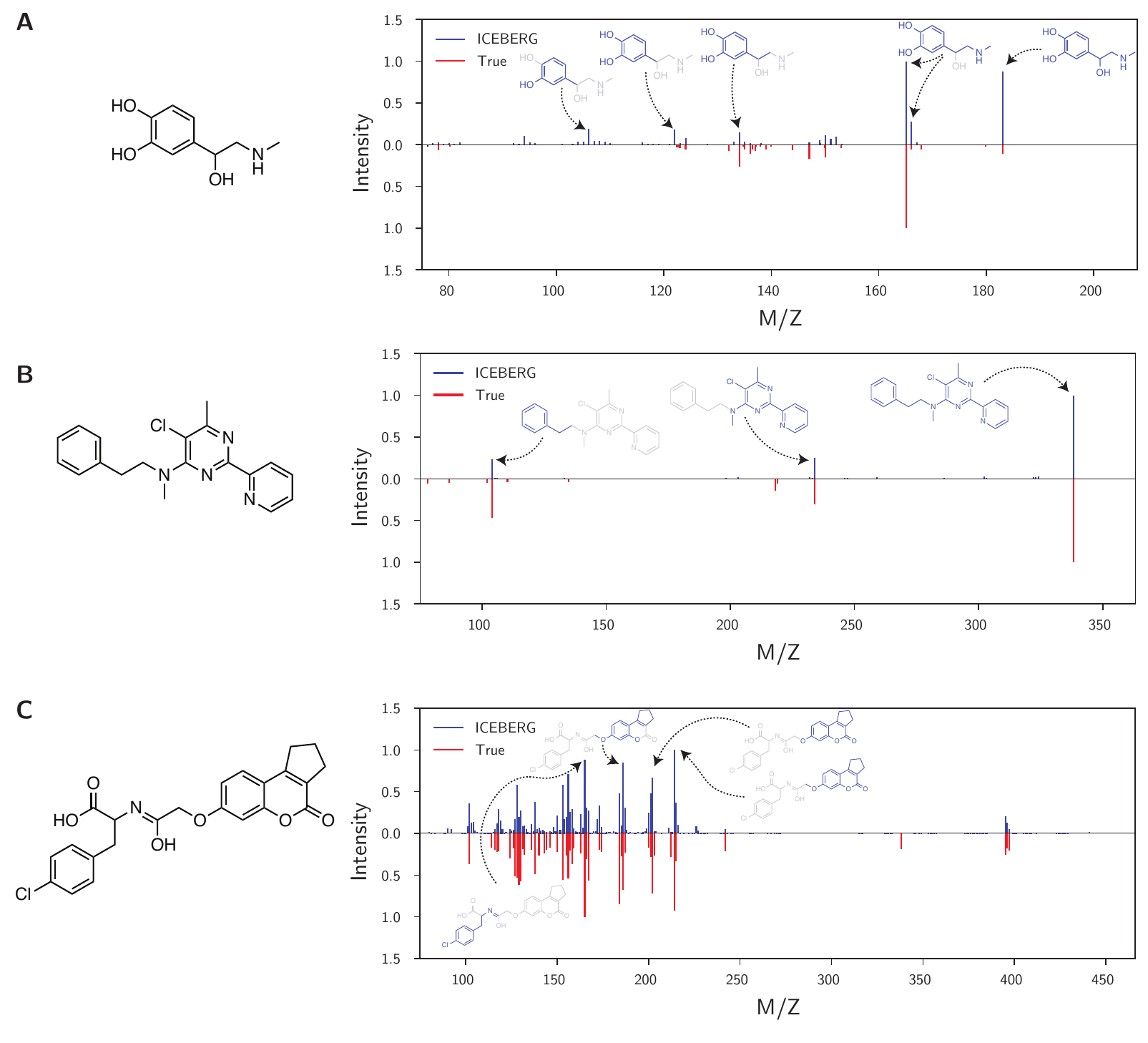}
  \caption{Examples of predicted spectra from \ourModel. Predictions are shown as generated by \ourModel trained on \gnpsData for select test set examples \texttt{GNPS:CCMSLIB00003137969} (\textbf{A}), \texttt{MoNA:001659} (\textbf{B}),  and \texttt{GNPS:CCMSLIB00000080524} (\textbf{C}).  The input molecular structures are shown (left); fragmentation spectra are plotted (right) with predictions (top, blue) and ground truth spectra (bottom, red). Molecular fragments are shown inset. Spectra are plotted with m/z shifted by the mass of the precursor adduct. All examples shown were not included in the model training set. }
  \label{fig:example_preds}
\end{figure}
\FloatBarrier

We evaluate \ourModel on its ability to accurately simulate positive ion mode mass spectra for both natural product like molecules and smaller organic molecules under 1,500 Da. Using the data cleaning pipeline from \cite{goldman_prefix-tree_2023}, we compile a public natural products dataset \gnpsData with 10,709 spectra (8,553 unique structures)  \cite{wang_sharing_2016, Goldman2023annotating, duhrkop_systematic_2021} as well as a gold standard chemical library \nistData with 35,129 spectra (24,403 unique structures) \cite{noauthor_tandem_nodate}. We note that \gnpsData was previously named `CANOPUS', renamed here to disambiguate the data from the tool CANOPUS \cite{duhrkop_systematic_2021}. Both datasets are split into structurally disjoint 90\%/10\% train-test splits, with 10\% of the training data reserved for model validation (see  \nameref{sec:methods_datasets}). We compare \ourModel against an expansive suite of contemporary and competitive methods \cite{allen_competitive_2015, hong20233DMolMS, murphy2023efficiently, wei_rapid_2019, zhu_using_2020, young_Massformer_2021, goldman_prefix-tree_2023}; all methods excluding \cfmModel are reimplemented, hyperparameter optimized, and trained on equivalent data splits, further described in the Supporting Information.

To measure performance, we calculate the average cosine similarity between each predicted spectrum and the true spectrum, as cosine similarity is widely used to cluster mass spectra in molecular networking \cite{nothias_feature-based_2020}. We find that \ourModel outperforms the next method \Massformer  on the natural product focused dataset (Figure \ref{fig:cosine_res}A). \ourModel achieves an average cosine similarity of 0.627, compared to \Massformer and \FixedVocab which each achieved a cosine similarity of 0.568---a 10\% improvement. 

Surprisingly, however, this boost in performance extends only marginally to the gold standard dataset, \nistData. \ourModel, while still outperforming binned spectrum prediction approaches (i.e., \texttt{NEIMS} \cite{wei_rapid_2019}) on this dataset, is nearly equivalent to \scarfModel  (0.727 v. 0.726) \cite{goldman_prefix-tree_2023}. Still, our model performs substantially better than \cfmModel and uses only a fraction of the computational resources (Figure \ref{fig:cosine_res}B). Unlike previous physically inspired models, because \ourModel only samples the most relevant fragments from chemical space, it requires just over 1 CPU second per spectrum.

We hypothesize that the discrepancy in performance improvement between \gnpsData and \nistData may be  partially explained by differences in the chemical spaces they cover. %
Many molecules within \gnpsData are natural products with more complicated chemical scaffolds. To characterize this, we analyzed the distributions for both the synthetic accessibility (SA) score \cite{ertl2009estimation, huang2022artificial} (Figure \ref{fig:cosine_res}C) and molecular weight (Figure \ref{fig:cosine_res}D), both proxies for molecular complexity. In concordance with our hypothesis, we find that SA scores and molecular weight are substantially higher on \gnpsData than \nistData: \gnpsData has an average SA score of 3.75, compared to 3.01 for \nistData; the datasets have average molecular weights of 413 Da and 317 Da respectively. %

\subsection{Model explanations of observed peaks are consistent with chemistry intuition}
\label{sec:spec_interpret}

\begin{figure}
  \centering
  \vspace{-1em}
  \includegraphics[width=0.5 \textwidth]{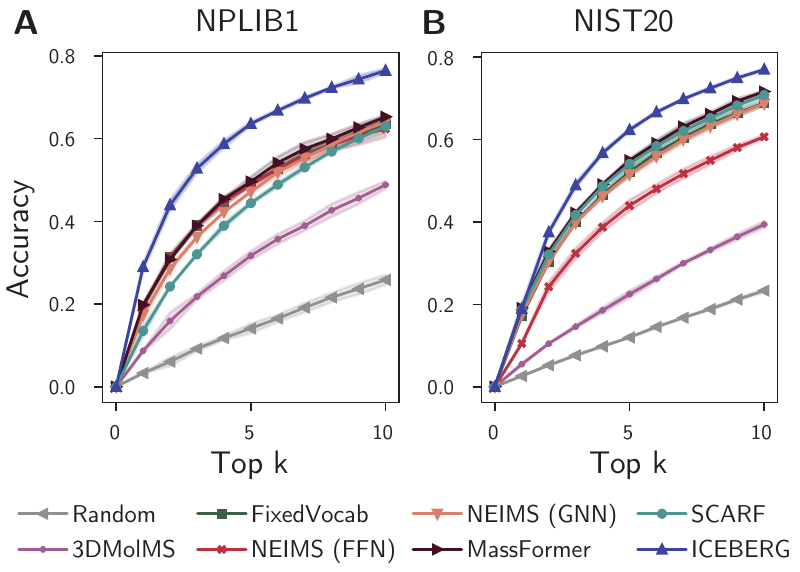}
  \caption{\ourModel enables improved spectrum retrieval over other methods on both \gnpsData (\textbf{A}) and \nistData (\textbf{B}) compared to other  spectrum prediction models. Top $k$ retrieval accuracy is computed by ranking a list of 49 additional candidates by putative cosine similarity as predicted by the model and determining the fraction of times that the true molecule is within the first $k$ entries. A region around each line is shaded using an upper and lower bound of 1.96 times the standard error of the mean across training seeds.}
\label{fig:retrieval}
\end{figure}

In addition to accurate predictions, a key benefit of simulating fragmentation events is that predictions are interpretable, even for  highly complex molecules. Each predicted peak from \ourModel is directly attributed to a fragment of the predicted molecule.

 By inspecting certain patterns and examples, we find expected broken bonds. Weaker bonds such as carbon-oxygen and carbon-nitrogen bonds tend to more reliably break, compared to carbon-carbon bonds and more complex ring breakages (Figure \ref{fig:example_preds}A). A second strength of using fragmentation-based models can be seen in
Figure \ref{fig:example_preds}B, where despite the heteroaromatic ring structures, our model is still able to correctly predict peak intensities by predicting a small number of carbon-nitrogen breakages.

Further alignment can be seen within the intensity prediction module. Because \ourModel predicts multiple intensities for each substructure corresponding to hydrogen shifts, 2 peaks can be present when a single bond breaks. In fragmentation example of  Figure \ref{fig:example_preds}A, the most intense peak is estimated at the mass shift of $-1\ch{H}$ from the original fragment, indicating that \ourModel correctly recognizes the hydroxyl group will likely leave as neutral $\ch{H2O}$ and result in a hydrogen rearrangement. We include additional, randomly selected prediction examples in the Supporting Information. %

\subsection{Fragmentation simulations lead to improved structural elucidation}
\label{sec:spec_annotation}

We next demonstrate that \ourModel improves the structural elucidation of unknown molecules using reference libraries of model-predicted spectra. %
We design a retrospective evaluation using our labeled data to resemble the prospective task of spectrum lookup within libraries. For each test spectrum, we extract up to $49$ ``decoy''  isomers  from PubChem \cite{kim_pubchem_2016} with the highest Tanimoto similarity to the true molecular structure. 
The consideration of up to 50 isomers mimics the realistic elucidation setting, as an unknown spectrum can yield clues regarding certain properties of its source molecule (e.g., computed using MIST \cite{Goldman2023annotating}, CSI:FingerID \cite{duhrkop_searching_2015}, or molecular networking \cite{nothias_feature-based_2020}), which narrows the chemical space of possible molecules to a smaller, more relevant set. %
We predict the fragmentation spectrum for each isomer and, for each model, we rank these possible matches by their spectral similarity to the spectrum of interest and compute how often the true molecule is found within the top \emph{k} ranked isomers for different values of \emph{k}.

We find that \ourModel improves upon the next best model by a margin of 9\% accuracy (a nearly 46\% relative improvement) in \emph{top 1} retrieval accuracy for the \gnpsData dataset (Figure \ref{fig:retrieval}A). Previous models with high spectrum prediction accuracies have struggled on this task due to their poor ability to differentiate structurally similar isomers  \cite{goldman_prefix-tree_2023}. Our  structure-based model appears to excel in retrieval and may have out-of-domain robustness beneficial to this task.

While the effect for top 1 retrieval accuracy on the \nistData dataset is not pronounced (i.e., tied with \Massformer), \ourModel outperforms the next best model by an absolute margin of over 5\% (a 7.5\% relative improvement) at top 10 accuracy (Figure \ref{fig:retrieval}B). These results underscore the real world utility of \ourModel to identify unknown molecules of interest.

To additionally evaluate the robustness of our model on the retrieval task in the real world, we further examine the accuracy of the highest performing methods to identify the correct molecule on the positive mode spectra from the recent Critical Assessment of Small Molecule Identification 2022 (CASMI22) competition~\cite{noauthor_critical_nodate}. Because CASMI22 is a natural products identification challenge, we test models trained on \gnpsData and test the four models with the highest top 1 retrieval accuracy on \gnpsData: \ourModel, \Massformer, \neimsFFN, and \FixedVocab. We find, once again, that \ourModel outperforms the other models tested on the retrieval task with 12.9\% accuracy compared to the next best \neimsFFN method that achieves a top 1 accuracy of 8.6\%. Absolute performance is low, but this is expected because: (1) we utilize a challenging PubChem~\cite{kim_pubchem_2019} retrieval library and (2) the retrieval accuracy for all entrants was relatively low; the entries submitted by the state of the art forward prediction model, \cfmModel~\cite{wang_cfm-id_2021}, only correctly predicted 2D structures for 26 of the 303 entries (8.6\%). Full results and comparison interpretations are discussed in the Supporting Information.

\subsection{Challenging, non-random data splits better explain retrieval performance}
\label{sec:spec_scaffold}

\begin{table}
\centering
\small
\caption{Comparing the accuracy of spectrum prediction on \nistData using random (easier) or scaffold (harder) split.}
    \begin{tabular}{lcc}
    \toprule
    {\nistData} &      Cosine sim.&   \\
    \cmidrule(r){2-3}
     & Random split &     Scaffold split\\
     \midrule
\cfmModel      &            0.412 &            0.411 \\
\MolMS     &            0.510 &            0.466 \\
\FixedVocab  &            0.704 &            0.658 \\
\neimsFFN &            0.617 &            0.546 \\
\neimsGNN &            0.694 &            0.643 \\
\Massformer  &            0.721 &            0.682 \\
\scarfModel       &            0.726 &            0.669 \\
    \midrule\midrule
\ourModel     &            \textbf{0.727} &            \textbf{0.699} \\
    \bottomrule
    \end{tabular}
\label{tab:scaffold}
\end{table}

The strong performance on the retrieval task, particularly for increasing values of $k$ on \nistData, suggests that \ourModel is able to generalize well to decoys not appearing in the training set and to account for how structural changes should affect fragmentation patterns. %
While encouraging, %
we observed only minor increases in cosine similarity accuracy when predicting spectra using \nistData (Figure \ref{fig:cosine_res}).

To try to explain this apparent discrepancy, we reevaluate prediction accuracy on %
a more challenging dataset split. We retrain all models on the \nistData utilizing a Murcko scaffold split of the data \cite{yang2019analyzing} with smaller scaffold clusters (i.e., more unique compounds) placed in the test set. This split enforces that molecules in the test set will be more distant and less similar to the training set, probing the ability of each model to generalize in a more stringent setting than our previous random split.

In the strict scaffold split evaluation, the improved accuracy of \ourModel over existing models is more apparent (Table \ref{tab:scaffold}). 
We find that \ourModel outperforms \Massformer and \scarfModel by 0.017 and 0.03-- 2\% and 4\% improvements respectively. %
These results suggest that, particularly for standard libraries with more homogeneous molecules, more challenging scaffold split evaluations may yield performance metrics that better correlate with performance on the structural elucidation problem (retrieval).

\section{Discussion}

We have proposed a physically-grounded mass spectrum prediction strategy we term \ourModel.  From a computational perspective, this integration of neural networks into fragmentation prediction is enabled by  (a) bootstrapping \MAGMA to construct fragmentation trees on which our model is trained, (b) posing the tree generation step as a sequential prediction over atoms, and (c) predicting multiple intensities at each generated fragment with a second module in order to account for  hydrogen rearrangements and isotopic peaks. 
By learning to generate fragmentation events, \ourModel is able to accurately predict mass spectra, yielding especially strong improvements for natural product molecules under evaluation settings of both spectrum prediction and retrieval. 

\ourModel  establishes new state of the art performance for these tasks, yet there are some caveats we wish to highlight. First, while we learn to generate molecular substructures to explain each peak, there are no guarantees that they are  the correct physical explanations given the number of potential equivalent-mass atom and bond rearrangements that could occur and our decision to train \ourModel \ourModelTwoShort to maximize a vectorized cosine similarity. Second, while we achieve increased accuracy, this comes at a higher computational cost of roughly 1 CPU second per molecule, nearly an order of magnitude more than other neural approaches like \scarfModel \cite{goldman_prefix-tree_2023}. 
Future work will consider more explicitly how to synergize fragment- and formula- prediction approaches to achieve higher accuracy and speed. In addition to model architecture modifications, we anticipate model accuracy improvements from modeling other covariates such as collision energy, adduct switching, instrument type, and even jointly modeling MS/MS with other analytical chemistry measurements such as FTIR \cite{fine2020spectral}. 

The discovery of unknown metabolites and molecules is rapidly expanding our knowledge of potential medical targets \cite{quinn_global_2020}, the effects of environmental toxins \cite{tian_2021_ubiquitous}, and the diversity of biosynthetically accessible chemical space \cite{doroghazi2014roadmap}. We envision exciting possibilities to apply our new model to expand the discovery of novel chemical matter from complex mixtures.

\section{Data and Software Availability}
All code to replicate experiments, train new models, and load pretrained models is available at \codeUrl{}. Pretrained models are available for download or as workflows through the GNPS2 platform~\cite{wang_sharing_2016} with an up-to-date link provided within the README file of our released GitHub code. %

\section{Funding}
This work was supported by the Machine Learning for Pharmaceutical Discovery and Synthesis consortium. S.G. was additionally supported by the MIT-Takeda Fellowship program. 

\section{Authors' contributions}
S.G. and J.L. jointly wrote the software. S.G. conducted experiments. S.G. and C.W.C. conceptualized the project and wrote the manuscript. C.W.C supervised the work. 

\begin{acknowledgement}
We thank John Bradshaw, Priyanka Raghavan, David Graff, Fanwang Meng, other members of the Coley Research Group, and Michael Murphy for helpful discussions, as well as Lucas Janson for feedback on earlier iterations of this idea. We thank Mingxun Wang for feedback and helpful suggestions regarding both the method and manuscript.

\end{acknowledgement}

\begin{suppinfo}
The Supporting Information provides an extended description of results, including exact values in tables, an additional evaluation on the CASMI22 dataset, additional spectrum prediction examples, further elaboration on the baselines we utilized, and hyperpameters tested. %
\end{suppinfo}

\bibliography{main}

\end{document}


\tableofcontents
\newpage

\begin{suppinfo}

\section{Extended results}

We include a complete set of all model results and metrics (i.e., Cosine similarity, coverage, validity, and time for 100 spectrum predictions) in Tables \ref{tab:spec_acc_nist} and \ref{tab:spec_acc_gnps}. Interestingly, while cosine similarity is higher for \ourModel on nearly all evaluations, the fraction of peaks in the ground truth spectrum explained by the predicted spectrum, \emph{coverage}, the fraction of peaks in the true spectrum explained by the predicted spectrum, does not increase. We posit that the more strict and rigid fragment-grounding causes our model to miss certain lower intensity peaks. In doing so, however, the model is able to maintain higher robustness to out-of-distribution molecules.

We note that not all predicted fragments contain valid molecular formula, as a very small fraction of predicted fragments fail the ring double bond equivalent (RDBE) molecular formula test \cite{pretsch2000structure} when hydrogen shifts are considered.

Full results for retrieval are shown for \nistData and \gnpsData and  in Tables 
\ref{tab:nist_spec_retrieval} and
\ref{tab:canopus_spec_retrieval} respectively.

\subsection{Independent data evaluation}
To further validate our models and inter-model comparison, we evaluated \ourModel on an independent challenge dataset. We selected the spectra utilized in the Critical Assessment of Small Molecule Identification 2022 (CASMI22) competition~\cite{noauthor_critical_nodate}, which contains a collection of 304 positive mode tandem mass spectra, as well as negative mode spectra that we do not consider herein. This challenge dataset was released after the compilation of all training data utilized for our models, so it reflects a realistic evaluation of our models trained on either \gnpsData or \nistData. 

We extract tandem mass spectra from the provided spectral data as described detailed in Goldman et al.~\cite{goldman_mist-cf_2023} and further filter the data by removing a single spectrum with a precursor mass >1,500 Da. To probe model retrieval accuracy on the remaining 303 challenge spectra, we compose a database of candidate molecules including the set of all isomers from the PubChem database~\cite{kim_pubchem_2019} for each challenge molecule. We note this assumes the correct identification of the precursor formula for each spectrum, which in actuality was achievable with >85\% accuracy in the contest. The resulting candidate database includes 268 unique formulae and 478,201 unique molecular candidates, an average of 1,784 candidates per formula. Most contestant entrants hand-crafted retrieval libraries for molecules they anticipated the CASMI competition organizers would utilize (natural products datasets, bioactive compounds, etc.). We emphasize that this decision may lead to our retrospective library selection conservatively underestimating our model's retrieval performance compared to other contest submissions. 

We evaluate our models on three different criteria: cosine similarity, top 1 retrieval accuracy, and the average Tanimoto similarity between the true molecule and our model's retrieved candidate. Because CASMI22 is organized by natural product scientists, we use models that we trained on the \gnpsData, rather than \nistData, and compare \ourModel against the subset of models with the highest retrieval performance for comparison. \ourModel is still the most performant method with a top 1 retrieval accuracy of 12.9\% (Table \ref{tab:casmi_retrieval}), outperforming other methods on cosine similarity and the two retrieval metrics. Curiously, all cosine similarities reported are far lower than in the standard test set evaluation, indicating that the CASMI22 dataset is challenging and that models may suffer from out-of-distribution bias. While comparing against other contest submissions is difficult due to the variance in the formula annotation accuracy, retrieval library construction, and manual post-processing, our retrieval accuracy is able to correctly identify >4\% more of the tested  spectra than \cfmModel, the most widely adopted spectrum prediction strategy.

\begin{figure}[t]
  \centering
  \includegraphics[width=0.8\textwidth]{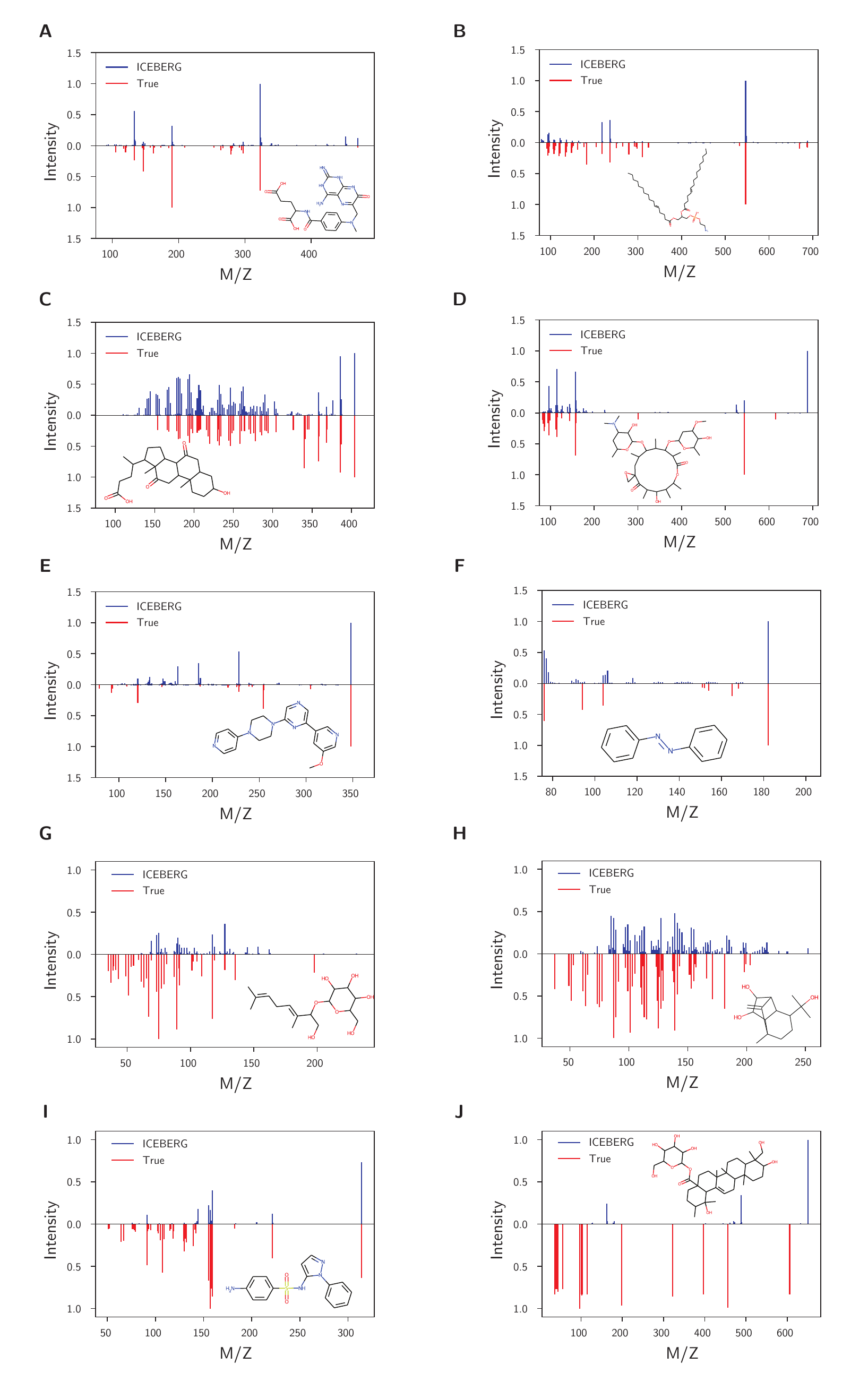}
  \caption{Additional \ourModel prediction examples. \ourModel (trained on \gnpsData) is utilized to make predictions on a randomly sampled set of 6 test set examples from \gnpsData (A-F) and also 4 randomly selected examples from CASMI22 (G-J). Predictions are shown in the top panel with true target spectra shown below. Test molecules are illustrated inset.}
  \label{fig:si_examples}
\end{figure}

\begin{table}
\centering
\caption{Independent CASMI 2022 dataset evaluation. A sub-selection of single models trained on the \gnpsData training dataset were re-evaluated on the CASMI22 dataset. Cosine similarity is computed on the 303 extracted CASMI22 positive mode spectra under 1,500 Da. Top 1 retrieval is computed using a PubChem retrieval dataset. The Top 1 Tanimoto similarity is computed between the retrieved molecule and the true molecule for each spectrum with a radius 2 Morgan fingerprint computed with RDKit~\cite{landrum2016rdkit}. The highest value in each column is typeset in bold. \\$\dagger$ The \cfmModel retrieval accuracy is computed directly from the \cfmModel CASMI22 submission which does not assume correct adducts but does customize the retrieval database to a smaller, more biased subset of candidates.}
\label{tab:casmi_retrieval}
\begin{tabular}{lccc}
\toprule
 & Cosine sim. & Top 1 retrieval accuracy & Top 1 Tanimoto similarity \\
\midrule
Random & \-- & 1.3\% & 0.196 \\
\midrule\midrule
\cfmModel & 0.248 & 8.6\%$^\dagger$ & \-- \\
\FixedVocab & 0.409 & 7.3\% & 0.324 \\
\neimsFFN & 0.361 & 8.6\% & 0.333 \\
\Massformer & 0.415 & 7.6\% & 0.317 \\
\midrule \midrule
\ourModel & \textbf{0.417} &\textbf{12.9\%} & \textbf{0.378} \\
\bottomrule
\end{tabular}
\end{table}

\begin{table}[ht]
\centering
\caption{Spectra prediction accuracy on \nistData, and \nistData (scaffold split) for \cfmModel \cite{allen_competitive_2015}, \MolMS \cite{hong20233DMolMS}, \FixedVocab \cite{murphy2023efficiently}, \neimsFFN \cite{wei_rapid_2019}, \neimsGNN \cite{zhu_using_2020}, \Massformer \cite{young_Massformer_2021}, \scarfModel \cite{goldman_prefix-tree_2023}, and \ourModel. Cosine similarity is calculated at a bin resolution of $0.1$ m/z; Coverage indicates the fraction of peaks in the ground truth spectrum explained by the predicted spectrum on average; Valid indicates the fraction of predicted peaks that can be explained as a subset of the precursor molecule's chemical formula (filtered to formulae with ring-double bond equivalents of greater or equal to zero); Time (s) indicates the number of seconds required to predict 100 random spectra from \nistData on a single CPU including the time to load the model. Values are shown $\pm$ 1.96 times the standard error of the mean across three random seeds for the same random split. Scaffold split experiments are only repeated once and shown without a corresponding interval. The best value in each column is typeset in bold.}
\label{tab:spec_acc_nist}
\resizebox{\textwidth}{!}{
\begin{tabular}{lrrrrrrr}
\toprule
Dataset & \multicolumn{3}{l}{\nistData (Random split)} & \multicolumn{3}{l}{\nistData (Scaffold split)} &  \\
\cmidrule(r){2-4} \cmidrule(r){5-7}
{Metric} &          Cosine sim. & Coverage & Valid &     Cosine sim. & Coverage &  Valid & Time (s) \\
\midrule
\cfmModel     &  $0.412 \pm 0.000$ &  $0.278 \pm 0.000$ &  \boldmath$1.00 \pm 0.000$ &  $0.411 $ &  $0.272 $ &  \boldmath$1.00 $ &  $1114.7$ \\
\MolMS     &  $0.510 \pm 0.001$ &  $0.734 \pm 0.002$ &  $0.94 \pm 0.002$ &  $0.466 $ &  $0.707 $ &  $0.97 $ &     \boldmath$3.5$ \\
\FixedVocab  &  $0.704 \pm 0.001$ &  $0.788 \pm 0.001$ &  \boldmath$1.00 \pm 0.000$ &  $0.658 $ &  $0.785 $ &  \boldmath$1.00 $ &     $5.5$ \\
\neimsFFN &  $0.617 \pm 0.001$ &  $0.746 \pm 0.002$ &  $0.95 \pm 0.001$ &  $0.546 $ &  $0.717 $ &  $0.96 $ &     $3.9$ \\
\neimsGNN &  $0.694 \pm 0.001$ &  $0.780 \pm 0.001$ &  $0.95 \pm 0.001$ &  $0.643 $ &  $0.766 $ &  $0.97 $ &     $4.9$ \\
\Massformer  &  $0.721 \pm 0.004$ &  $0.790 \pm 0.006$ &  $0.93 \pm 0.006$ &  $0.682 $ &  $0.789 $ &  $0.96 $ &     $7.7$ \\
\scarfModel       &  $0.726 \pm 0.002$ &  \boldmath$0.807 \pm 0.001$ &  \boldmath$1.00 \pm 0.000$ &  $0.669 $ &  \boldmath$0.807 $ &  \boldmath$1.00 $ &    $21.1$ \\
\midrule\midrule
\ourModel     &  \boldmath$0.727 \pm 0.002$ &  $0.754 \pm 0.002$ &  \boldmath$1.00 \pm 0.000$ &  \boldmath$0.699 $ &  $0.771 $ &  \boldmath$1.00 $ &    $82.2$ \\
\bottomrule
\end{tabular}}
\end{table}

\begin{table}[ht]
\centering
\caption{Spectra prediction accuracy on \gnpsData for \cfmModel \cite{allen_competitive_2015}, \MolMS \cite{hong20233DMolMS}, \FixedVocab \cite{murphy2023efficiently}, \neimsFFN \cite{wei_rapid_2019}, \neimsGNN \cite{zhu_using_2020}, \Massformer \cite{young_Massformer_2021}, \scarfModel \cite{goldman_prefix-tree_2023}, and \ourModel. Cosine similarity is calculated at a bin resolution of $0.1$ m/z; Coverage indicates the fraction of peaks in the ground truth spectrum explained by the predicted spectrum on average; Valid indicates the fraction of predicted peaks that can be explained as a subset of the precursor molecule's chemical formula (filtered to formulae with ring-double bond equivalents of greater or equal to zero). Values are shown $\pm$ 1.96 times the standard error of the mean across three random seeds on a single test split. The best value in each column is typeset in bold. }
\label{tab:spec_acc_gnps}
\resizebox{0.8\textwidth}{!}{
\begin{tabular}{lrrr}
\toprule
Dataset & \multicolumn{3}{l}{\gnpsData}  \\
\cmidrule(r){2-4}
{Metric} &  Cosine sim. & Coverage &  Valid  \\
\midrule
\cfmModel      &    $0.377 \pm 0.000$ &  $0.235 \pm 0.000$ &  \boldmath$1.00 \pm 0.000$ \\
\MolMS     &    $0.394 \pm 0.004$ &  $0.507 \pm 0.001$ &  $0.92 \pm 0.001$ \\
\FixedVocab  &    $0.568 \pm 0.003$ &  $0.563 \pm 0.002$ &  \boldmath$1.00 \pm 0.000$ \\
\neimsFFN &    $0.491 \pm 0.003$ &  $0.524 \pm 0.002$ &  $0.95 \pm 0.001$ \\
\neimsGNN &    $0.521 \pm 0.003$ &  $0.547 \pm 0.005$ &  $0.94 \pm 0.001$ \\
\Massformer  &    $0.568 \pm 0.004$ &  \boldmath$0.573 \pm 0.009$ &  $0.95 \pm 0.003$ \\
\scarfModel       &    $0.536 \pm 0.013$ &  $0.552 \pm 0.016$ &  \boldmath$1.00 \pm 0.000$ \\
\midrule\midrule
\ourModel     &    \boldmath$0.627 \pm 0.002$ &  $0.549 \pm 0.003$ &  \boldmath$1.00 \pm 0.000$ \\
\bottomrule
\end{tabular}}
\end{table}

\begin{table}[ht]
\centering
\caption{\nistData spectra retrieval top k accuracy for different values of $k$, $\pm$ 1.96 times the standard error of the mean across three random seeds on the same test set.}
\label{tab:nist_spec_retrieval}
\resizebox{\textwidth}{!}{
\begin{tabular}{lrrrrrrrrrr}
\toprule
Top k &                  1  &                  2  &                  3  &                  4  &                  5  &                  8  &                  10 \\
\midrule
Random      &  $ 0.026 \pm 0.001$ &  $ 0.052 \pm 0.001$ &  $ 0.076 \pm 0.002$ &  $ 0.098 \pm 0.001$ &  $ 0.120 \pm 0.001$ &  $ 0.189 \pm 0.003$ &  $ 0.233 \pm 0.004$ \\
\MolMS     &  $ 0.055 \pm 0.003$ &  $ 0.105 \pm 0.000$ &  $ 0.146 \pm 0.005$ &  $ 0.185 \pm 0.007$ &  $ 0.225 \pm 0.009$ &  $ 0.332 \pm 0.005$ &  $ 0.394 \pm 0.008$ \\
\FixedVocab  &  $ 0.172 \pm 0.004$ &  $ 0.304 \pm 0.004$ &  $ 0.399 \pm 0.002$ &  $ 0.466 \pm 0.007$ &  $ 0.522 \pm 0.012$ &  $ 0.638 \pm 0.009$ &  $ 0.688 \pm 0.006$ \\
\neimsFFN &  $ 0.105 \pm 0.003$ &  $ 0.243 \pm 0.012$ &  $ 0.324 \pm 0.013$ &  $ 0.387 \pm 0.011$ &  $ 0.440 \pm 0.014$ &  $ 0.549 \pm 0.010$ &  $ 0.607 \pm 0.005$ \\
\neimsGNN &  $ 0.175 \pm 0.005$ &  $ 0.305 \pm 0.003$ &  $ 0.398 \pm 0.002$ &  $ 0.462 \pm 0.004$ &  $ 0.515 \pm 0.005$ &  $ 0.632 \pm 0.007$ &  $ 0.687 \pm 0.005$ \\
\Massformer  &  \boldmath$ 0.191 \pm 0.008$ &  $ 0.328 \pm 0.006$ &  $ 0.422 \pm 0.003$ &  $ 0.491 \pm 0.002$ &  $ 0.550 \pm 0.005$ &  $ 0.662 \pm 0.005$ &  $ 0.716 \pm 0.003$ \\
\scarfModel       &  $ 0.187 \pm 0.008$ &  $ 0.321 \pm 0.011$ &  $ 0.417 \pm 0.007$ &  $ 0.486 \pm 0.008$ &  $ 0.541 \pm 0.009$ &  $ 0.652 \pm 0.008$ &  $ 0.708 \pm 0.009$ \\
\midrule\midrule
\ourModel     &  $ 0.189 \pm 0.012$ &  \boldmath$ 0.375 \pm 0.005$ &  \boldmath$ 0.489 \pm 0.007$ &  \boldmath$ 0.567 \pm 0.005$ &  \boldmath$ 0.623 \pm 0.004$ &  \boldmath$ 0.725 \pm 0.003$ &  \boldmath$ 0.770 \pm 0.002$ \\
\bottomrule
\end{tabular}}
\end{table}

\begin{table}[ht]
\centering
\caption{\gnpsData spectra retrieval top k accuracy for different values of $k$, $\pm$ 1.96 times the standard error of the mean across three random seeds on the same test set.}
\label{tab:canopus_spec_retrieval}
\resizebox{\textwidth}{!}{
\begin{tabular}{lrrrrrrrrrr}
\toprule
Top k &                  1  &                  2  &                  3  &                  4  &                  5  &                  8  &                  10 \\
\midrule
Random      &  $ 0.033 \pm 0.002$ &  $ 0.061 \pm 0.010$ &  $ 0.092 \pm 0.007$ &  $ 0.118 \pm 0.005$ &  $ 0.141 \pm 0.012$ &  $ 0.216 \pm 0.012$ &  $ 0.258 \pm 0.012$ \\
\MolMS     &  $ 0.087 \pm 0.003$ &  $ 0.159 \pm 0.020$ &  $ 0.218 \pm 0.008$ &  $ 0.268 \pm 0.012$ &  $ 0.317 \pm 0.011$ &  $ 0.427 \pm 0.016$ &  $ 0.488 \pm 0.010$ \\
\FixedVocab  &  $ 0.193 \pm 0.007$ &  $ 0.314 \pm 0.008$ &  $ 0.390 \pm 0.005$ &  $ 0.448 \pm 0.010$ &  $ 0.492 \pm 0.003$ &  $ 0.587 \pm 0.010$ &  $ 0.635 \pm 0.011$ \\
\neimsFFN &  $ 0.195 \pm 0.005$ &  $ 0.313 \pm 0.005$ &  $ 0.388 \pm 0.006$ &  $ 0.447 \pm 0.012$ &  $ 0.488 \pm 0.003$ &  $ 0.585 \pm 0.014$ &  $ 0.624 \pm 0.020$ \\
\neimsGNN &  $ 0.174 \pm 0.014$ &  $ 0.285 \pm 0.008$ &  $ 0.362 \pm 0.004$ &  $ 0.422 \pm 0.001$ &  $ 0.471 \pm 0.003$ &  $ 0.586 \pm 0.013$ &  $ 0.640 \pm 0.010$ \\
\Massformer  &  $ 0.198 \pm 0.003$ &  $ 0.308 \pm 0.004$ &  $ 0.389 \pm 0.001$ &  $ 0.454 \pm 0.004$ &  $ 0.496 \pm 0.012$ &  $ 0.599 \pm 0.012$ &  $ 0.653 \pm 0.005$ \\
\scarfModel       &  $ 0.135 \pm 0.014$ &  $ 0.242 \pm 0.001$ &  $ 0.320 \pm 0.003$ &  $ 0.389 \pm 0.008$ &  $ 0.444 \pm 0.004$ &  $ 0.569 \pm 0.003$ &  $ 0.630 \pm 0.015$ \\
\midrule\midrule
\ourModel     &  \boldmath$ 0.290 \pm 0.008$ &  \boldmath$ 0.439 \pm 0.013$ &  \boldmath$ 0.528 \pm 0.010$ &  \boldmath$ 0.587 \pm 0.009$ &  \boldmath$ 0.636 \pm 0.004$ &  \boldmath$ 0.723 \pm 0.001$ &  \boldmath$ 0.764 \pm 0.005$ \\
\bottomrule
\end{tabular}}
\end{table}

\FloatBarrier
\section{Graph neural network details}
\label{sec:mol_enc}
\ourModel relies upon graph neural network embeddings of the molecule $\gnn{\mol}$. Given the widespread use and descriptions of such models, we refer the reader to \citeauthor{Li2015-fc} for a description of the gated graph neural networks we employ. We utilize the DGL library \cite{Wang2019-ga} to implement and featurize molecular graphs. Because our fragmentation method relies upon iteratively removing atoms, we often have molecular fragments with incomplete valence shells that would not be parsed by RDKit \cite{landrum2016rdkit}. As such, we opt for a more minimal set of atom and bond features described in Table~\ref{table:atomFeats} .

\begin{table}[ht]
  \centering
  \caption{Dataset details.}
  \begin{tabular}{llll}
    \toprule
    Name         & \# Spectra & \# Molecules & Description                            \\
    \midrule
    \nistData & 35,129 & 24,403 & Standards library \\
    \gnpsData & 10,709 & 8,533 & Natural products from GNPS \cite{wang_sharing_2016} \\
    \bottomrule
\end{tabular}
\label{table:dataset_details}
\end{table}

\begin{table}[ht]
  \centering
  \caption{Graph neural network (GNN) atom features.}
  \begin{tabular}{ll}
    \toprule
    Name         & Description                            \\
    \midrule
    Element type & one-hot encoding of the element type                    \\
    Hydrogen number & one-hot encoding the number of hydrogens on each atom                    \\
    Adduct type  & one-hot encoding of the ionization adduct                    \\
    Random walk embed steps & positional encodings of the nodes computed using DGL \\
    \bottomrule
\end{tabular}
\label{table:atomFeats}
\end{table}

\section{Baselines}
\label{sec:baselines}
We compare \ourModel to three classes of spectrum prediction models: \emph{binned prediction} models that encode a molecule and output a single fixed length discretized spectrum vector, \emph{fragmentation prediction} models such as \ourModel that directly learn to fragment and break the input molecule, and \emph{formula prediction} models that predict chemical formula peaks and corresponding intensities. We briefly describe each model used below,  emphasizing the novelty of their contribution. 

For all baselines that utilize 2D graph neural networks that are not pretrained (i.e,. \neimsGNN), SCARF, and \FixedVocab), we utilize a common graph neural network base architecture consisting of gated graph neural network message passing steps as in \ourModel (see  \nameref{sec:mol_enc}) \cite{Li2015-fc}. Baseline GNNs are provided with a wide range of atom features including element type, the degree of each atom, hybridization type, formal charge, a binary flag denoting whether the atom is in a ring-system, atomic mass as a floating point value, a one hot encoding of the chiral tag, the type of the atom, and a random walk embedding step. All models tested are provided the same set of experimental covariates: a onehot vector corresponding to the adduct type. For graph models, this is concatened as an atom level feature, whereas in other models, it is concatenated to the intermediate hidden representation.

\paragraph{Binned prediction} 
\begin{enumerate}
    \item \neimsFFN: \citeauthor{wei_rapid_2019} encode a fingerprint representation of a molecule with a feed forward neural network and predict a fixed dimensional discrete binned output vector. In addition to predicting a fragment vector, \citeauthor{wei_rapid_2019} also predict the mass differences from the precursor ion as an independent vector and combine the two outputs with a learned, weighted sum. Originally trained on EI mass spectrometry, we replicate this approach on our datasets of ESI mass spectra. 
    
    Key hyperparameters are whether or not to use mass differences in the output prediction, the number of feed forward layers, hidden size, and dropout between hidden layers. 
    \item \neimsGNN): \citeauthor{zhu_using_2020} extend the NEIMS model to utilize graph neural networks instead of feed forward networks applied to fingerprints. We utilize our default graph neural network layer scheme described above, emphasizing that the novelty of this contribution is to switch to a 2D graph neural network encoder from fingerprints.  In such a case, we also concatenate model covariates (i.e., adduct one hot) as additional atom level features to the GNN).

    Key hyperparameters are whether or not to use mass differences in the output prediction, the number of GNN layers, the number of random walk embedding steps (i.e., positional encodings of the nodes computed using DGL), and dropout between hidden layers. 
    \item \Massformer: \citeauthor{young_Massformer_2021} further extend the \neimsGNN) approach using a \emph{pretrained} graph neural network model, the Graphormer \cite{ying_transformers_2021}. Rather than train the model entirely from scratch to make a mass and mass difference prediction vector, \Massformer encodes molecules using a large, 48M parameter model previously pretrained. Our \Massformer implementation replaces the GNN with the Graphormer model. Covariates (i.e., adduct type one hots) are concatenated after the Graphormer encoding. We utilize the same scheme of \citeauthor{young_Massformer_2021} to use the pretrained Graphomer Version 2, reinitialize all 12 intermediate Graphormer layers, reinitialize the layer norm, and keep only the learned token embeddings. We allow this model to predict mass differences as in NEIMS by default.

    Key hyperparameters are the number of feed forward layers after the Graphormer encoding, the dropout between these layers, and the hidden dimension size.
    \item \MolMS: \citeauthor{hong20233DMolMS} replace the 2D graph neural network architecture with a 3D point cloud graph neural network to predict binned spectra. However, they do not predict mass differences as in \Massformer or NEIMS, substantially reducing model performance in our reimplementation setting. 
    
    Key hyperparameters are the hidden size of their encoder, number of layers in the 3D neural network, number of top layers after the 3D neural network, and the number of neighbors to utilize for message passing in 3D space.
\end{enumerate}  

\paragraph{Fragmentation prediction} 
\begin{enumerate}
    \item \cfmModel: \citeauthor{wang_cfm-id_2021} parametrize a Markov model of fragmentation by first combinatorially fragmenting a molecule of interest then subsequently estimating transition probabilities across the fragmentation graph to predict peak intensities. We utilize the pretrained version 4 available for public release, rather than retraining the model.
\end{enumerate}  

\paragraph{Formula prediction} 
\begin{enumerate}
    \item \FixedVocab: \citeauthor{murphy2023efficiently} recently introduced \graffMS, a model that predicts spectra at the level of chemical formula by pre-defining a list of $10,000$ common chemical formula and chemical formula differences from the precursor. They encode the input molecule with a graph neural network and predict intensities at each predefined chemical formula. Due to unavailability of source code, we define a variation of their model we term \FixedVocab that predicts intensities at a library of the most popular formula fragments and formula losses in the training dataset. We utilize a cosine similarity loss function to train the model. We revise the name of this model to delineate the differences in loss function and consideration of isotopes and adducts between our implementations.

    Key hyperparameters are the hidden size of the model, number of graph neural network layers, random walk embedding steps, graph pooling, and the size of the fixed-length formula vocabulary.

    An earlier version of this work under-reported the accuracy of this model, as the implementation did not allow for the prediction of precursor compound masses.
    \item \scarfModel: \citeauthor{goldman_prefix-tree_2023} use a similar strategy to \citeauthor{murphy2023efficiently}, but rather than use a fixed vocabulary, they autoregressively generate formula candidates as a prefix tree first with the \scarfThread model. This smaller vocabulary size conditioned on the input molecule enables them to deploy a second neural network model \scarfWeave to explicitly encode the candidate formulae with a neural network alongside the molecule in a second modeling step to predict intensities.

    We refer the reader to \cite{goldman_prefix-tree_2023} for a complete description of hyperparameters, as this model is used without changes from the original work. 
\end{enumerate}  

\section{Hyperparameters}
\label{sec:hyps}

To fairly compare the various methods, we apply a rigorous hyperparameter optimization scheme. We use RayTune \cite{liaw2018tune} with Optuna \cite{akiba2019optuna} and an ASHAScheduler to identify hyperparameters from a grid set of options. All models were allotted 50 different hyperoptimization trials on a $10,000$ spectra subset of \nistData. Hyperparameter descriptions for \ourModel are provided in Table \ref{tab:sup-hyper-params}. Selected baseline and \ourModel hyperparameters are defined in Tables \ref{tab:sup-hyper-baselines} and \ref{tab:sup-hyper-ours} respectively. When possible, we recycle parameters as selected in \citeauthor{goldman_prefix-tree_2023}, as the hyperparameter optimization scheme is equivalent.

\begin{table}[ht]
  \centering
  \caption{Hyperparameter descriptions for \ourModel.}
  \resizebox{\textwidth}{!}{
  \begin{tabular}{lll}
    \toprule
    Name         &  Model (\ourModelOneShort or \ourModelTwoShort) & Description                            \\
    \midrule
    learning rate & both & optimizer learning rate                     \\
    learning rate decay ($5,000$ steps) & both & step-wise learning rate decay every 5,000 model weight update steps                    \\
    dropout & both & model dropout applied in-between linear hidden layers               \\
    hidden size & both & number of hidden layers               \\
    gnn layers & both & number of graph neural network layers to encode molecules and fragments    \\
    mlp layers & both & number of feed forward layers to encode concatenated representations               \\
    transformer layers & \ourModelTwoShort & number of set transformer attention layers after mlp encoding    \\
    batch size & both & number of spectra to include in each training batch    \\
    weight decay & both & optimizer weight decay    \\
    random walk embed steps & \ourModelOneShort & number of random walk embedding steps to for graph neural network atom features \\
    graph pooling & both & how to combine atom features into a single representation\\
    bin size & \ourModelTwoShort & binned spectrum resolution spacing from $0$ Da to $1,500$ Da \\
    \bottomrule
\end{tabular}}
\label{tab:sup-hyper-params}
\end{table}

\begin{table}[ht]
\centering
\caption{\ourModel \ourModelOneShort and \ourModelTwoShort hyperparameter grid and selected values.}
\label{tab:sup-hyper-ours}
\begin{tabular}{@{}llll@{}}
\toprule
\textbf{Model} & \textbf{Parameter}                               & \textbf{Grid}               & \textbf{Value} \\ \midrule
\ourModel \ourModelOneShort   & learning rate                                    & $[1e-4,1e-3]$               & $0.00099$      \\
               & learning rate decay ($5,000$ steps)                               & $[0.7, 1.0]$              & 0.7214            \\
               & dropout                                          & $\{0.1,0.2,0.3\}$           & $0.2$          \\
               & hidden size                                 & $\{128, 256, 512\}$        & $512$          \\
               & mlp layers                                      & \--                 & $1$            \\
               & gnn layers                                    & $[1,6]$                 & $6$            \\
               & batch size                                      & $\{8, 16, 32, 64\}$                 & $32$            \\
               & weight decay                                      & $\{0, 1e-6, 1e-7\}$                 & $0$            \\
               & random walk embed steps                                       & [0,20]                         & 14           \\
               & graph pooling                                       & \{mean, attention\}                         & mean           \\

\midrule
\ourModel \ourModelTwoShort   & learning rate                                    & $[1e-4,1e-3]$               & $0.00074$      \\
               & learning rate decay ($5,000$ steps)                               & $[0.7, 1.0]$              & 0.825            \\
               & dropout                                          & $\{0.1,0.2,0.3\}$           & $0.1$          \\
               & hidden size                                 & $\{128, 256, 512\}$        & $256$          \\
               & mlp layers                                     & $[0,3]$                 & $1$            \\
               & gnn layers                                      & $[1,6]$                 & $4$            \\
               & transformer layers                                     & $[0,3]$                 & $3$            \\
               & batch size                                      & $\{8, 16, 32,\}$                 & $32$            \\
               & weight decay                                      & $\{0, 1e-6, 1e-7\}$                 & $1e-7$            \\
               & bin size                                     & \--                         & $0.1$           \\
               & graph pooling                                       & \{mean, attention\}                         & mean           \\

\midrule
\end{tabular}
\end{table}

\FloatBarrier

\begingroup
\small
\begin{singlespace}

\setlength{\tabcolsep}{4pt} %
\begin{centering}
\begin{longtable}{@{}llll@{}}
\caption{Baseline hyperparameters.}
\label{tab:sup-hyper-baselines}\\
\toprule
\textbf{Model} & \textbf{Parameter}                               & \textbf{Grid}               & \textbf{Value}
\endfirsthead
\multicolumn{4}{c}%
{{\bfseries \tablename\ \thetable{} -- continued from previous page}} \\
\midrule \textbf{Model} & \textbf{Parameter}                               & \textbf{Grid}               & \textbf{Value}\\\midrule
\endhead
\midrule
\neimsFFN    & learning rate                                    & $[1e-4,1e-3]$               & $0.00087$      \\
               & learning rate decay ($5,000$ steps)                               & $[0.7, 1.0]$              & 0.722            \\
               & dropout                                          & $\{0.0, 0.1,0.2,0.3\}$           & $0.0$          \\
               & hidden size, $d$                                 & $\{64,128,256,512\}$        & $512$          \\
               & layers, $l$                                      & $\{1,2,3\}$                 & $2$            \\
               & batch size                                      & $\{16, 32, 64, 128\}$                 & $128$            \\
               & weight decay                                      & $\{0, 1e-6, 1e-7\}$                 & $0$            \\
               & use differences                                    & \{True, False\}                 & True            \\
               & bin size                                         & \--                         & $0.1$           \\

\midrule
\neimsGNN)            & learning rate                                    & $[1e-4,1e-3]$               & $0.00052$      \\
               & learning rate decay ($5,000$ steps)                               & $[0.7, 1.0]$              & 0.767            \\
               & dropout                                          & $\{0.0, 0.1,0.2,0.3\}$           & $0.0$          \\
               & hidden size, $d$                                 & $\{64,128,256, 512\}$        & $512$          \\
               & layers, $l$                                      & $[1,6]$                 & $4$            \\
               & batch size                                      & $\{16, 32, 64\}$                 & $64$            \\
               & weight decay                                      & $\{0, 1e-6, 1e-7\}$                 & $1e-7$            \\
               & use differences                                      & \{True, False\}                 & True            \\
               & bin size                                         & \--                         & $0.1$           \\
               & random walk embed steps                                      & [0,20]                         & 19           \\
               & graph pooling                                       & \{mean, attention\}                         & mean           \\
\midrule
\MolMS            & learning rate                                    & $[1e-4,1e-3]$               & $0.00074$      \\
               & learning rate decay ($5,000$ steps)                               & $[0.7, 1.0]$              & 0.86            \\
               & dropout                                          & $\{0.0, 0.1,0.2,0.3\}$           & $0.3$          \\
               & hidden size, $d$                                 & $\{64,128,256, 512\}$        & $256$          \\
               & layers, $l$                                      & $[1,6]$                 & $2$            \\
               & top layers                                      & $[1,3]$                 & $2$            \\
               & neighbors, $k$                                      & $[3,6]$                 & $5$            \\
               & batch size                                      & $\{16, 32, 64\}$                 & $16$            \\
               & weight decay                                      & $\{0, 1e-6, 1e-7\}$                 & $1e-6$            \\
               & bin size                                         & \--                         & $0.1$           \\
\midrule
\Massformer            & learning rate                                    & $[1e-4,1e-3]$               & $0.00014$      \\
               & learning rate decay ($5,000$ steps)                               & $[0.7, 1.0]$              & 0.85            \\
               & feed forward dropout                                          & $\{0.0, 0.1,0.2,0.3, 0.4, 0.5\}$           & $0.1$          \\
               & feed forward hidden size, $d$                                 & $\{64,128,256, 512, 1024\}$        & $1024$          \\
               & feed forward layers, $l$                                      & $[1,6]$                 & $1$            \\
               & batch size                                      & $\{32, 64, 128\}$                 & $128$            \\
               & weight decay                                      & $\{0, 1e-6, 1e-7\}$                 & $1e-7$            \\
               & bin size                                         & \--                         & $0.1$           \\
\midrule

\FixedVocab            & learning rate                                    & $[1e-4,1e-3]$               & $0.00018$      \\
               & learning rate decay ($5,000$ steps)                               & $[0.7, 1.0]$              & 0.92            \\
               & dropout                                          & $\{0.0, 0.1,0.2,0.3\}$           & $0.3$          \\
               & hidden size, $d$                                 & $\{64,128,256, 512\}$        & $512$          \\
               & layers, $l$                                      & $[1,6]$                 & $6$            \\
               & batch size                                      & $\{16, 32, 64\}$                 & $64$            \\
               & weight decay                                      & $\{0, 1e-6, 1e-7\}$                 & $1e-6$            \\
               & bin size                                         & \--                         & $0.1$           \\                
               & random walk embed steps                                      & [0,20]                         & 11           \\
               & graph pooling                                       & \{mean, attention\}                         & mean   \\
              & formula library size                                       & \{1000, 5000, 10000, 25000, 50000\}                         & 5000   \\
\midrule
\scarfThread & learning rate                                    & $[1e-4,1e-3]$               & $0.000577$      \\
               & learning rate decay ($5,000$ steps)                               & $[0.7, 1.0]$              & 0.894            \\
               & dropout                                          & $\{0.0, 0.1,0.2,0.3\}$           & $0.3$          \\
               & hidden size, $d$                                 & $\{128, 256, 512\}$        & $512$          \\
               & mlp layers, $l_1$                                      & $[1,3]$                 & $2$            \\
               & gnn layers, $l_2$                                      & $[1,6]$                 & $4$            \\
               & batch size                                      & $\{8, 16, 32, 64\}$                 & $16$            \\
               & weight decay                                      & $\{0, 1e-6, 1e-7\}$                 & $1e-6$            \\
               & use differences                                        & \{True, False\}                 & True            \\
               & random walk embed steps                                         & [0,20]                         & 20           \\
               & graph pooling                                       & \{mean, attention\}                         & mean           \\

\midrule
\scarfWeave   & learning rate                                    & $[1e-4,1e-3]$               & $0.00031$      \\
               & learning rate decay ($5,000$ steps)                               & $[0.7, 1.0]$              & 0.962            \\
               & dropout                                          & $\{0.0, 0.1,0.2,0.3\}$           & $0.2$          \\
               & hidden size, $d$                                 & $\{128, 256, 512\}$        & $512$          \\
               & mlp layers, $l_1$                                     & $[1,3]$                 & $2$            \\
               & gnn layers, $l_2$                                     & $[1,6]$                 & $3$            \\
               & transformer layers, $l_3$                                     & $[0,3]$                 & $2$            \\
               & batch size                                      & $\{4, 8, 16, 32, 64\}$                 & $32$            \\
               & weight decay                                      & $\{0, 1e-6, 1e-7\}$                 & $0$            \\
               & bin size                                      & \--                         & $0.1$           \\
               & random walk embed steps                                        & [0,20]                         & 7           \\
               & graph pooling                                       & \{mean, attention\}                         & attention           \\

\midrule

\end{longtable}
\end{centering}
\end{singlespace}
\endgroup  

\end{suppinfo}

\bibliography{main}